\documentclass[longauth]{aa}

\usepackage{graphicx}
\usepackage{graphics}
\usepackage{txfonts}
\usepackage{epsfig}
\usepackage{natbib}
\usepackage{hyperref}
\usepackage[]{color}

\begin{document}

\title{Detection of orbital motions near the last stable circular orbit of the massive black hole SgrA*}

\titlerunning{Detection of Orbital Motions Near the ISCO of the SMBH SgrA*}
\subtitle{}

\author{GRAVITY Collaboration\thanks{GRAVITY is developed
    in a collaboration by the Max Planck Institute for
    Extraterrestrial Physics, LESIA of Paris Observatory/Universit\'e PSL/CNRS/Sorbonne Universit\'e/Univ.
Paris Diderot/Sorbonne Paris Cit\'e, IPAG of Universit\'e Grenoble Alpes/CNRS,
     the Max Planck Institute for Astronomy, the University of
    Cologne, the CENTRA - Centro de Astrof\'isica e Gravita\c{c}\~{a}o, and
    the European Southern Observatory. \newline Corresponding authors:
    O.~Pfuhl (\href{mailto:pfuhl@mpe.mpg.de}{pfuhl@mpe.mpg.de}), J.~Dexter  (\href{mailto:jdexter@mpe.mpg.de}{jdexter@mpe.mpg.de}), T.~Paumard  (\href{mailto:thibaut.paumard@obspm.fr}{thibaut.paumard@obspm.fr}) \& R.~Genzel (\href{mailto:genzel@mpe.mpg.de}{genzel@mpe.mpg.de})}:
R.~Abuter\inst{8}
\and A.~Amorim\inst{6,14}
\and M.~Baub\"ock\inst{1}
\and J.P.~Berger\inst{5}
\and H.~Bonnet\inst{8}
\and W.~Brandner\inst{3}
\and Y.~Cl\'{e}net\inst{2}
\and V.~Coud\'e~du~Foresto\inst{2}
\and P.T.~de~Zeeuw\inst{10,1}
\and C.~Deen\inst{1}
\and J.~Dexter\inst{1}
\and G.~Duvert\inst{5}
\and A.~Eckart\inst{4,13}
\and F.~Eisenhauer\inst{1}
\and N.M.~Förster~Schreiber\inst{1} 
\and P.~Garcia\inst{7,9,14}
\and F.~Gao\inst{1}
\and E.~Gendron\inst{2}
\and R.~Genzel\inst{1,11}
\and S.~Gillessen\inst{1}
\and P.~Guajardo\inst{9}
\and M.~Habibi\inst{1}
\and X.~Haubois\inst{9}
\and Th.~Henning\inst{3}
\and S.~Hippler\inst{3}
\and M.~Horrobin\inst{4}
\and A.~Huber\inst{3}
\and A.~Jim\'enez-Rosales\inst{1}
\and L.~Jocou\inst{5}
\and P.~Kervella\inst{2}
\and S.~Lacour\inst{2,1}
\and V.~Lapeyr\`ere\inst{2}
\and B.~Lazareff\inst{5}
\and J.-B.~Le~Bouquin\inst{5}
\and P.~L\'ena\inst{2}
\and M.~Lippa\inst{1}
\and T.~Ott\inst{1}
\and J.~Panduro\inst{3}
\and T.~Paumard\inst{2}
\and K.~Perraut\inst{5}
\and G.~Perrin\inst{2}
\and O.~Pfuhl\inst{1}
\and P.M.~Plewa\inst{1}
\and S.~Rabien\inst{1}
\and G.~Rodr\'iguez-Coira\inst{2}
\and G.~Rousset\inst{2}
\and A.~Sternberg\inst{12,15}
\and O.~Straub\inst{2}
\and C.~Straubmeier\inst{4}
\and E.~Sturm\inst{1}
\and L.J.~Tacconi\inst{1}
\and F.~Vincent\inst{2}
\and S.~von~Fellenberg\inst{1}
\and I.~Waisberg\inst{1}
\and F.~Widmann\inst{1}
\and E.~Wieprecht\inst{1}
\and E.~Wiezorrek\inst{1} 
\and J.~Woillez\inst{8}
\and S.~Yazici\inst{1,4}
}

\institute{
Max Planck Institute for Extraterrestrial Physics(MPE), Giessenbachstr.1, 85748 Garching, Germany
\and LESIA, Observatoire de Paris, Universit\'e PSL, CNRS, Sorbonne Universit\'e, Univ. Paris Diderot, Sorbonne Paris Cit\'e, 5 place Jules Janssen, 92195 Meudon, France
\and Max Planck Institute for Astronomy, K\"onigstuhl 17, 69117,
Heidelberg, Germany
\and $1^{\rm st}$ Institute of Physics, University of Cologne,
Z\"ulpicher Straße 77, 50937 Cologne, Germany
\and Univ. Grenoble Alpes, CNRS, IPAG, 38000 Grenoble, France
\and Universidade de Lisboa - Faculdade de Ci\^{e}ncias, Campo Grande, 1749-016 Lisboa, Portugal
\and Faculdade de Engenharia, Universidade do Porto, rua Dr. Roberto Frias, 4200-465 Porto, Portugal
\and European Southern Observatory, Karl-Schwarzschild-Str. 2, 85748
Garching, Germany
\and European Southern Observatory, Casilla 19001, Santiago 19, Chile
\and Sterrewacht Leiden, Leiden University, Postbus 9513, 2300 RA
Leiden, The Netherlands
\and Departments of Physics and Astronomy, Le Conte Hall, University
of California, Berkeley, CA 94720, USA
\and School of Physics and Astronomy, Tel Aviv University, Tel Aviv 69978, Israel
\and Max Planck Institute for Radio Astronomy, Auf dem H\"ugel 69, 53121
Bonn, Germany
\and CENTRA - Centro de Astrof\'isica e Gravita\c{c}\~{a}o, IST, Universidade de Lisboa, 1049-001 Lisboa, Portugal
\and Center for Computational Astrophysics, Flatiron Institute, 162 5th Ave., New York, NY, 10010, USA
}

\date{Oct. 18, 2018 }

\abstract{We report the detection of continuous positional and polarization changes of the compact source SgrA* in high states (`flares') of its variable near-infrared emission with the near-infrared GRAVITY-Very Large Telescope Interferometer (VLTI) 
beam-combining instrument. In three prominent bright flares, the position centroids exhibit clockwise looped motion on the sky, on scales of typically 150 micro-arcseconds over a few tens of minutes, corresponding to about 30\% the speed of light. At the same time, the flares exhibit continuous rotation of the polarization angle, with about the same $\rm45 (\pm15)$-minute period as that of the centroid motions. Modelling with relativistic ray tracing shows that these findings are all consistent with a near face-on, circular orbit of a compact polarized `hot spot' of infrared synchrotron emission at approximately six to ten times the gravitational radius of a black hole of 4 million solar masses. This corresponds to the region just outside the innermost, stable, prograde circular orbit (ISCO) of a Schwarzschild-Kerr black hole, or near the retrograde ISCO of a highly spun-up Kerr hole. The polarization signature is consistent with orbital motion in a strong poloidal magnetic field.
}

\keywords{Galactic centre -- general relativity -- black holes}

\maketitle

\section{Introduction}
\label{sec:intro}
 
The compact source SgrA* at the centre of the Milky Way harbours a concentration of 4.14 million solar masses, plausibly a massive black hole \citep{genzel2010,ghez2008}. SgrA* exhibits steady and continuously variable, non-thermal emission across the electromagnetic spectrum \citep{genzel2010,baganoff2001,trippe2007,eckart2008,yusef-zadeh2008,do2009,dodds-eden2009,shahzamanian2015,ponti2017,witzel2018}. Intercontinental microwave interferometry and polarized infrared(IR)/X-ray variability on 10-30 minute timescales suggest that this emission comes from highly relativistic electrons in a hot, magnetized accretion disk/torus of $\sim$10 light minutes in diameter, plus perhaps a jet, just outside the innermost stable circular orbit (ISCO) of the putative massive black hole \citep{witzel2018,johnson2018,doeleman2008,yuan2004,markoff2001}. The exploration of this innermost, relativistic accretion region with high-resolution imaging techniques promises important and fundamental information for physics and astronomy, including new stringent tests of the massive black hole paradigm. \\
We have been observing the Galactic centre and SgrA* with the GRAVITY instrument \citep{GRAVITY2017,GRAVITY2018a,eisenhauer2011,eisenhauer2008,paumard2008} during multiple campaigns in 2017/2018\footnote{ESO Telescopes at the La Silla Paranal Observatory  programme IDs 099.B-0162, 0100.B-0731, 0101.B-0195, and 0101.B-0576.}, with the aim of testing general relativity (GR) and the massive black hole paradigm in the closest massive black hole candidate. \cite{GRAVITY2018a} have already reported in this journal on a high-quality measurement of the gravitational redshift in the orbit of the star S2 going through its peri-approach at  2800\,$R_g$ from SgrA* ($R_g= GM_{\bullet}/c^2=6.1\times10^{11}\, \rm cm$, or 5$\,\rm \mu as$)
in May 2018. Another main goal of our observations is to search for orbital motions of `hot spots' of relativistic gas in the innermost accretion zone around the black hole's ISCO \citep{broderick2005,broderick2006,hamaus2009}. \\
Such hot spots have been proposed to originate from magnetic shocks or re-connection events in the innermost accretion zone \citep{eckart2008,zamaninasab2010,dexter2013,chan2009,dodds-eden2010,ponti2017} leading to local acceleration of electrons to relativistic $\rm \gamma$-factors of $10^{3...6}$, sufficient to generate the variable IR (and X-ray) emission, in analogy to solar flares \citep{dodds-eden2010,lin2001,ponti2017}. This is the subject of the current paper.

\section{Observations}
\label{sec:observations}
The GRAVITY instrument combines the four 8\,m telescopes of the European Southern Observatory (ESO)-Very Large Telescope (VLT) interferometrically for 3 milli-arcsec (mas) resolution imaging and $\sim$20-70 microarcsec ($\rm\mu as$) astrometry in the K-band (2.2\,$\rm \mu m$) continuum. For details of the instrument and the data analysis and positional extraction we refer to \cite{GRAVITY2017} and Appendix A. Briefly, the light of the four telescopes is extracted into mono-mode fibres for two positions on the sky and then interfered in the beam combiner for all six baselines of the interferometer. One fibre is placed on the bright ($K_s=10$) star IRS16C about 1" N-E of SgrA*, and the other is on SgrA*, plus the orbiting star S2. In 2018 these two sources were conveniently separated by only 14-20\,mas, which is less than the fibre diameter (50\,mas) and the diffraction beam of an 8m UT at 2.2\,$\rm \mu m$ (56\,mas). This allowed precise, continuous measurements of the positional separation vector between the K-band continuum emission of SgrA* and S2, $\delta \vec{r}(t)_{\rm SgrA*-S2}$, while the fringes were detected and stabilised with the second fibre on IRS16C. Since IR flares are polarized, we recorded the Stokes component Q, or Q and U.\\
The astrometric position precision of GRAVITY measurements in good atmospheric conditions depends on the magnitude of SgrA* (relative to S2, which has $K_s=14$) and the integration time. While SgrA* is detected in $\sim$90\% of our science frames in 2018, its median magnitude is $\sim$17, which requires a co-addition of more than an hour to reach an rms astrometric precision of $\rm \sigma_{astromet}{\sim}30\,\mu arcsec$ ($\rm \mu as$). However, during bright `states' (henceforth referred to as `{\bf flares}', with $ K_s\le15$, whose probability of occurrence is $<10^{-2}$, \citealt{dodds-eden2011,witzel2018}) this positional precision can be reached in integrations of  2-10 minutes. Three additional corrections to the raw astrometry are required. The orbital motion of the reference (S2) over an hour observation is about 10 micro-arcseconds, which is subtracted from $\delta \vec{r}(t)_{\rm SgrA*-S2}$ . Since the near-IR spectral energy distribution (SED) of SgrA* is significantly redder than that of S2, its flux weighted wavelength across the instrumental bandpass is larger than that of S2, and  $\delta \vec{r}(t)_{\rm SgrA*-S2}$ has to be corrected by 1.001. Likewise, differential refraction results in corrections of 10-20\,$\rm\mu as$. With these aspects taken into account,  it is possible to detect the expected motions of a compact, hot spot orbiting a black hole of  4 million solar masses near the ISCO, which for a non-rotating black hole corresponds to an orbital diameter of 60\,$\rm \mu as$ and an orbital period of 31 minutes.

\section{Results}
\label{sec:results}

\subsection{Emission centroid motions}
\label{subsec:centroids}

\begin{figure*}[ht]
\includegraphics[width=18cm]{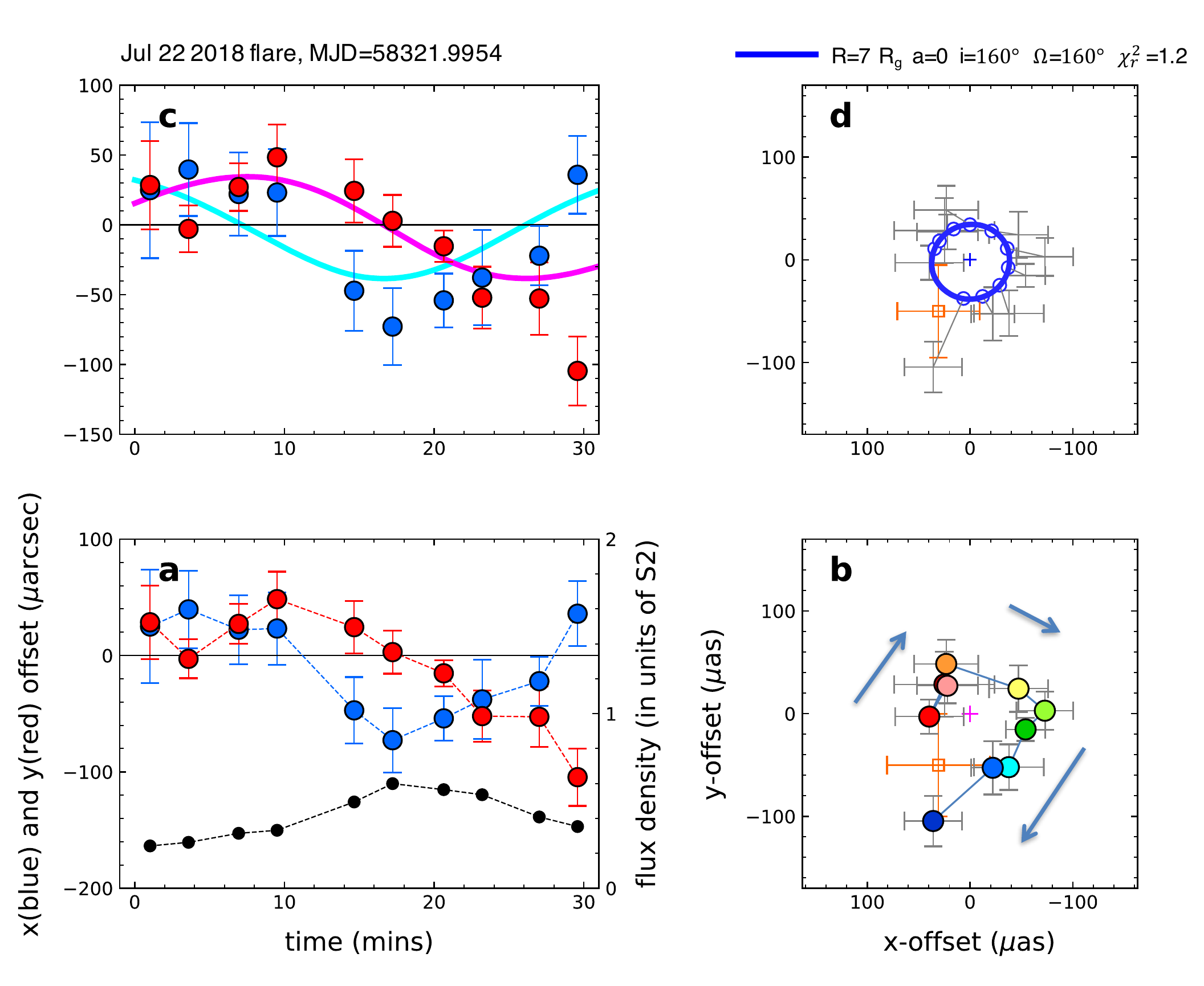}
\caption{Bottom left (a): Time evolution of the east-west (east positive, blue) and north-south (red) position offset of the July 22$^{\rm }$ flare (MJD=58321.9954) centroids from their medians, as well as the flux density evolution (right y-axis, black) in units of the flux of S2 (14.0 mag). Error bars are 1$\sigma$. For this purpose the total intensity was computed from the sum of the two polarization directions. The points represent the average of the `Pfuhl' and `Waisberg' analyses (see Appendix \ref{sec:AppB}). Bottom right (b): Projected orbit of the flare centroid on the sky (colour ranging from brown to dark blue as a qualitative marker of time through the 30-minute observation, relative to their medians (small black cross) and after removal of the S2 motion and differential refraction between S2 and SgrA*). The orange square and 1$\sigma$ uncertainty is the long-term astrometric position of the mass centre of the S2 orbit (approximately the orbital centroid, although shifts between apparent and true centroids can be introduced by lensing, relativistic beaming, and azimuthal shearing of an initially compact `hot spot'). Top left (c) and top right (d): Comparison of the data of the bottom two panels with a realization of a simple hot spot model in the Schwarzschild metric, including light bending, lensing, time dilation and other effects of GR and/or special relativity (SR), computed from the {\bf  NERO} relativistic ray tracing code (Baub\"ock et al. in prep.). Similar results were obtained with the {\bf GYOTO} code \citep{vincent2011,grould2016}. The purple and cyan continuous curves in (c) show the same orbit in $x(t)$ and $y(t)$, compared to the data in blue and red. The continuous blue curve in (d) denotes a hot spot on a circular orbit with $R=1.17\times R({\rm ISCO}, a=0, M=4.14\times10^6 M_{\odot})$, seen at inclination $160^{\circ}$ (clockwise on the sky, as for the data in (d)) and with the line of nodes at $\Omega=160^{\circ}$ (${\chi_{r}}^{2}=1.2$). Open blue circles and grey bars connect the data points to their locations on the best fit orbit. }
\label{fig:Fig1}
\end{figure*}

We observed two bright flares with a peak approaching the flux of S2 on July $\rm 22^{}$ and $\rm July ~28, 2018,^{}$ as  well as a fainter flare ($0.3-0.5 \times $ S(S2))  on May $\rm27,^{} 2018^{}$. These flares lasted for 30 to 90 minutes. Figure \ref{fig:Fig1} (a) shows for the July $\rm 22^{}$ flare the RA (blue) and Dec (red) positional offsets (from their medians) of the SgrA* emission centroid as a function of time, along with the flux evolution in units of S2's flux ($K_S=14$, or 15 mJy, black). At all times, K-band emission from  SgrA* is spatially unresolved. We detect significant and continuous positional changes of the emission centroid in both coordinates, of $\sim$120 $\rm \mu as$ over $\sim$30 minutes. This corresponds to $\sim$0.3 times the speed of light. The motions appear to trace out 50-70\% of a closed, clockwise loop (Figure  \ref{fig:Fig1} (b,d)). In RA/Dec the data can be fitted well by sine-curves of the same amplitude, characteristic for a closed circular orbit observed near face-on (Figure \ref{fig:Fig1} (c)). Four independent analyses of these data with different software by different GRAVITY team members give comparable and consistent results on these motions. Figure \ref{fig:Fig1} uses the average of the `Pfuhl (P)' and `Waisberg (W)' analyses, which are also shown separately in Figure \ref{fig:FigB1}  (Appendix \ref{sec:AppB}). The position of SgrA*'s mass centroid derived from the S2 orbit \citep{GRAVITY2018a} is consistent with the centroid of the flare orbit, to within the $\rm \pm50\,\mu as$ uncertainties (orange square and cross in Figure  \ref{fig:Fig1} (b) and (d)).

\begin{table}
      \caption[]{{\bf GYOTO} fit parameters for hot spot orbits of the three flares. Analysis `Pfuhl' (P), `Waisberg' (W). `LESIA' (L) and average of P and W (P+W).}
      \label{tab:a1}
      \begin{tabular}{lllllllll}
            \hline
            \noalign{\smallskip}
            flare & analysis & $a$ & $R$ & $i$ & $\Omega$ & ${\chi_r}^2$ $(N_{\rm dof})$  \\
             &  &   & $(R_g)$ &$(^\circ)$ & $(^\circ)$ &  \\
            \noalign{\smallskip}
            \hline
            \noalign{\smallskip}
           $\rm Jul~22^{nd}$ & P & $0 $ & $7.3$ & 164 & 118 & 1.1 (18)\\
              &   & &  &  &  &   \\
               & W & $0 $ & $7.2$ & 151 & 122 & 1.9 (14) \\
              &   & &  &  &  &   \\ 
              & P+W & $0 $ & $8.0$ & 149 & 115 & 1.3 (14)  \\
              &   & &  &  & &   \\ 
              & P & $-1 $ & $7.7$ & 151 & 109 & 1.07 (18) \\
              &   & &  &  &  &   \\ 
              & L & $ 0 $ & $7.1$ & 167 & 185 & 1.80 (16) \\
              &   & &  &  &  &   \\     
              & L  & $ -1 $ & $7.4$ & 157 & 164 & 1.65 (16) \\
              &   & &  &  &  &   \\               
              $\rm Jul~28^{}$ & P & $0 $ & $9.1$ & 164 & 103 & 4.5 (30)\\
              &   & &  &  &  &   \\
              & P & $-1 $ & $9.1$ & 152 & 110 & 4.3  (30) \\
              &   &  &  &  &  &   \\
              $\rm May~27^{}$ & P & $0 $ & $8.3$ & 179 & 131 & 2.7 (16)\\
              &   & &  &  &  &   \\
              & P & $-1 $ & $7.1$ & 163 & 127 & 2.1  (16) \\
              &   & &  & &  &  \\
            \noalign{\smallskip}
            \hline
            \hline
         \end{tabular}
\end{table}

\begin{table}
      \caption[]{{\bf NERO} fit parameters for hot spot orbits of the three flares. For description see caption of Table \ref{tab:a1}. }
      \label{tab:a2}
      \begin{tabular}{llllllllll}
            \hline
            \noalign{\smallskip}
            flare & analysis & $a$ & $R$ & $i$ & $\Omega$  & ${\chi_r}^2$ $(N_{\rm dof})$  \\
             &  &   & $(R_g)$ &$(^\circ)$ & $(^\circ)$ &  \\
            \noalign{\smallskip}
            \hline
            \noalign{\smallskip}
           $\rm Jul~22^{nd}$ & P & $0 $ & $8$ & 145 & 126 & 1.6 (16)\\
              &   & &  &  &  &  & \\
               & W & $0 $ & $7$ & 140 & 160 & 1.6 (16) \\
              &   & &  &  & &  & \\ 
              & P+W & $0 $ & $7$ & 160 & 160  & 1.2 (16) \\
              &   & &  &  &  & \\ 
              & L & $0 $ & $7$ & 160 & 160  & 1.6 (16) \\
              &   & &  &  &  & \\ 
              $\rm Jul~28^{th}$ & P & $0 $ & $9$ & 135 & 137 & 4.5 (30)\\
              &   & &  &  &  & \\
              $\rm May~27^{th}$ & P & $0 $ & $7$ & 170 & 34 & 2.1 (16)\\
              &   & &  &  &  & \\
            \noalign{\smallskip}
            \hline
           \hline
         \end{tabular}
\end{table}
We discuss in Appendix\,\ref{sec:AppB} measurements of a second, similarly bright flare on July 28$^{\rm }$, 2018, (MJD=58328.0841) and a fainter flare on May 27$^{\rm }$, 2018 (MJD=58266.3420). The data in Figure \ref{fig:FigB4} broadly exhibit the same properties as those in Figure \ref{fig:Fig1}. Again the emissions for these two flares appear to follow incomplete clockwise loops of total maximum to minimum amplitude $\sim$110 to 140 $\rm\mu$as over a time period of $\sim$40-70 minutes, and again the loop centroids are consistent within the uncertainties with SgrA*'s mass position. Overall the data quality is somewhat poorer than in Figure \ref{fig:Fig1}, owing to the less favourable atmospheric conditions and lower fluxes. The short lifetimes observed for all three best flares are expected in models of  hot spots \citep{broderick2005,broderick2006,hamaus2009} since the differential rotation in an accretion disk would in a single orbital period shear out an initially compact gas cloud into an elongated arc. \\
The apparent orbits of the flares, in addition to their common clockwise motion, appear to be relatively face on (Tables \ref{tab:a1} and \ref{tab:a2}). Our best fits of the data in Figure \ref{fig:Fig1} with the {\bf NERO} and {\bf GYOTO} relativistic ray tracing codes (Baub\"ock et al. in prep, \citealt{vincent2011}) including light bending, lensing, time dilation etc.) yield $R\approx7\pm0.5 \,R_g$, a line of nodes at $\Omega\approx115^{\circ}-160^{\circ}$ and an inclination $i\approx160^{\circ}\pm10^{\circ}$, with an orbital period of $P=40\pm8$ minutes, for orbits in an angular momentum parameter $a=0$ (Schwarzschild) space time (Tables \ref{tab:a1} and \ref{tab:a2}). See also Appendix \ref{sec:AppB}, Figure \ref{fig:FigB2}, and Table \ref{tab:a2}. Tables \ref{tab:a1} and \ref{tab:a2} summarise the fit results for all three flares obtained with two independent analyses of the data and two independent relativistic fit codes. {\bf All three flares can in principle be accounted for by a single orbit model.} Interestingly this common orbit shows a similar orientation and angular momentum direction as the clockwise stellar disk and the G2 object ($\Omega\approx99^{\circ}$, $i\approx129^{\circ}$, \citealt{bartko2009,genzel2010,gillessen2012,pfuhl2015,plewa2017}). Early estimates of the orientation of the hot gas in the innermost accretion zone based on radio data suggest a similar orientation as well ($\Omega\sim 128..157^{\circ} $  E. of N. (note $180^{\circ}$ degeneracy), yet higher inclination $i\sim50..68^{\circ}$, \citealt{dexter2010,broderick2011,wang2013}).\\
Figures \ref{fig:Fig1} and \ref{fig:FigB4} show that flux distribution in all three flares is not strongly peaked on any one side of the line of nodes, as would be expected from relativistic beaming for an inclined orbit. We show in Appendix \ref{subsec:doppler} that inclinations of $>155^{\circ}$ (or equivalently $<25^{\circ}$) would reduce the expected impact of relativistic beaming of radiation to below a factor of 1.5 at one to two times ISCO, which is consistent with our data and the relativistic orbit modelling (Tables \ref{tab:a1} and \ref{tab:a2}), since we see little evidence for rapid or even superluminal centroid motions due to strong light bending or multiple images (Figure \ref{fig:FigB2}, \citealt{broderick2005,broderick2006,hamaus2009}). In cases of such low inclinations, the observed flux variations do not primarily reflect variations in beaming and lensing during an orbit but are more determined by the balance of local heating and cooling of the relativistic electrons. The synchrotron cooling time is $\tau_{synchr}=15\times \left(\frac{B}{20\,\rm G}\right)^{-1.5} \left( \frac{\lambda}{2.2\, \rm \mu m}\right)^{0.5}$ minutes \citep{gillessen2006}.
Theoretical estimates of the magnetic field in the innermost accretion zone around SgrA* vary between 5 and 100 G \citep{yuan2003,moscibrodzka2009,dodds-eden2010,dexter2010,ponti2017}. If the hot spot is created by magnetic reconnection, the local magnetic energy density is tapped for heating of the electrons, and as a result $B^2/8\pi$ drops considerably, lengthening the cooling time \citep{dodds-eden2010,ponti2017}. It is thus uncertain whether the observed flare durations are determined by the cooling time and/or the shearing of the hot spot due to differential rotation.

\subsection{Rotation of polarization angles}
\label{subsec:polarization}
\begin{figure*}[ht]
\includegraphics[width=17cm]{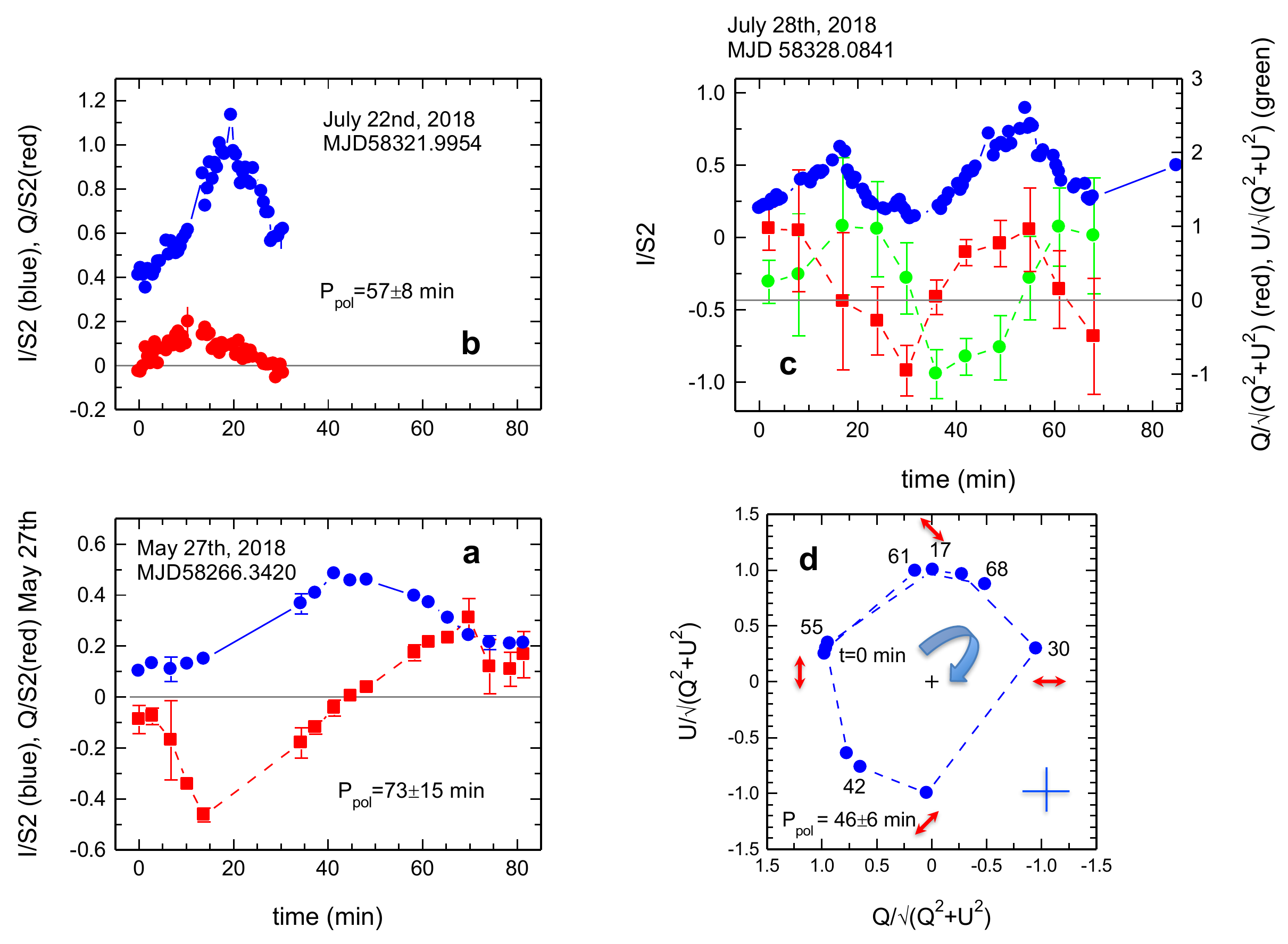}
\caption{Bottom left (a): Total flux I/S2 (blue, relative to S2) and Q/I flux ratio as a function of time for the May 27$^{\rm }$ flare. Top left (b): As in (a) but for the July 22 flare.  Top right (c): I/S2 (blue), $Q/(Q^2+U^2)^{1/2}$ (red) and $U/(Q^2+U^2)^{1/2}$  (green) evolution during the  July 28$^{\rm }$ flare. Bottom right (d): Evolution of the July 28$^{\rm }$ flare in the plane of normalized Stokes parameters $Q/(Q^2+U^2)^{1/2}$ (horizontal) and $U/(Q^2+U^2)^{1/2}$ (vertical). The red arrows denote the polarization directions on sky. The blue cross in the lower right denotes a typical error bar.}
\label{fig:Fig2}
\end{figure*}

In addition to the astrometric motions we observe systematic and continuous variations of the position angle of the polarization of the IR synchrotron emission. Figure \ref{fig:Fig2} (a) shows a full-turn change of Q/I projection over $P_{\rm pol} \approx73\pm15$ minutes during the May 27$^{\rm }$ flare, and Figure \ref{fig:Fig2} (b) indicates half a turn of polarization change in $26\pm8$ minutes during the July 22$^{\rm }$ flare (in the latter two cases we only recorded one polarization). Figure \ref{fig:Fig2} (c,d) shows the evolution of the $U/(Q^2+U^2)^{1/2}$ ($\pm45^{\circ}$) versus $Q/(Q^2+U^2)^{1/2}$ ($0/90^{\circ}$) normalized Stokes components during the July 28 flare. The data appear to trace out a loop structure in time, with a period of $P_{\rm pol}\approx 48\pm6$ minutes, again suggestive of orbital motion of an ordered field/polarization structure (see Appendix \ref{sec:AppD}).  We note that significant swings of polarization angle have previously been observed in NACO polarimetry of a few SgrA* flares \citep{trippe2007,zamaninasab2010,shahzamanian2015}, albeit not as impressive and complete as in Figure \ref{fig:Fig2}. We show in Appendix \ref{sec:AppD} that such smooth polarization swings (with $P_{\rm pol}\sim P_{\rm orbit}$) can be accounted for if the magnetic field axis is orthogonal to the orbital axis, as in a poloidal field configuration, and if the orbiting hot spot is observed at low inclination. {\bf The Q-U loops are caused by light-bending effects.} If the field instead were dominated by a toroidal configuration in the plane of the motions, one would expect $P_{\rm pol}=0.5\times P_{\rm orbit}$.

\section{Discussion - Evidence for orbital motion in the deep relativistic zone}
\label{sec:discussion}

\begin{figure}
\includegraphics[width=9cm]{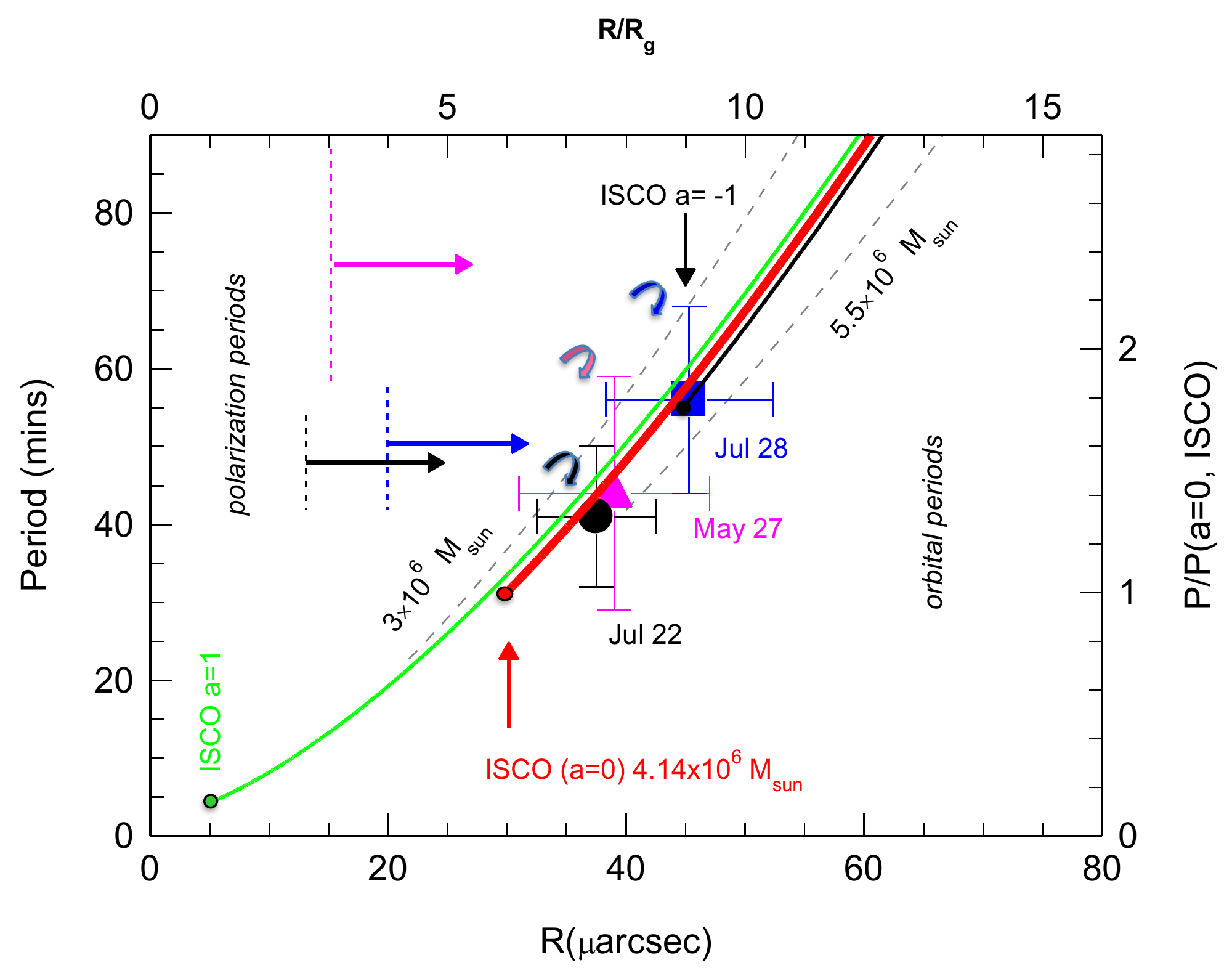}
\caption{Radius (R) - orbital period ($P_{\rm orb}$) relations for flare loops on July 22$^{\rm }$ (black), July 28$^{\rm }$ (blue), and May 27$^{\rm }$ (pink) compared to circular orbits around a black hole of 4.14 (solid lines), 3, and 5.5 (dashed grey lines) million solar masses. The solid red curve is for orbits around a spin 0 (Schwarzschild) black hole \citep{schwarzschild1916,bardeen1972}. The solid green and black lines are for prograde and retrograde orbits around a maximum (spin parameter $|a|=1$) Kerr black hole \citep{kerr1963}. Small green, red, and black circles mark the innermost stable circular orbits (ISCO) in these cases \citep{bardeen1972}. For the three flare loops, the filled black circle, blue square, and open pink triangles mark the (R, P) estimates from the best Schwarzschild ($a=0$) orbit fits with the relativistic ray tracing codes (Figures \ref{fig:Fig1}, \ref{fig:FigB3},  \ref{fig:FigB4}, Tables \ref{tab:a1} and \ref{tab:a2}). In principle all three flares can be accounted for by the {\bf same} Schwarzschild $R\approx6-10\, R_g$ circular orbit with a period of $\sim$45 minutes. Interestingly the data can also be fit by a retrograde orbit on ISCO for an extreme Kerr hole ($a=-1$). The periods $P_{\rm pol}$ for the full swings of the polarization orientation of the July 22$^{\rm }$, July 28$^{\rm }$ and May 27$^{\rm }$ flares are given as right pointing black, blue, and pink arrows (with their rms uncertainties marked as dotted vertical lines; see Figure \ref{fig:Fig2}). For a polarized hot spot in a poloidal field the polarization rotation periods are equal to the spot orbital period, in agreement with our data.}
\label{fig:Fig3}
\end{figure}

Figure \ref{fig:Fig3} summarises our constraints on orbit radii and periods of the flares, interpreted as orbiting compact hot spots, and on the expected orbit kinematics around a massive black hole. As discussed above, in this simple hot spot scenario, {\bf the centroid motions as well as the polarization rotation in all three observed flares (July 22$^{\rm }$, July 28$^{\rm }$, May 27, 2018$^{\rm  }$) are broadly consistent with similar circular orbits of a polarized hot spot.} To obtain more quantitative estimates, the impact of orbit broadening due to astrometric errors, due to the incomplete orbital coverage and the effects of GR and SR need to be included. When these effects are included in circular orbit models (Figures \ref{fig:Fig1}, \ref{fig:FigB3}, \ref{fig:FigB4}), our analysis shows that the flare centroid motions and polarization swings of all three flares are plausibly consistent with the same circular orbit of $R=6-10 \,R_g$ and $P=33-65$ minutes (Tables \ref{tab:a1} and \ref{tab:a2}). This common orbit is at $\rm 1.17~ (\pm0.3)$ ISCO for a low spin ($a\approx0$, `Schwarzschild') black hole (for 4 million solar masses, \citealt{schwarzschild1916,bardeen1972}). These data constrain the mass of the Schwarzschild hole to be the same to within $\pm30$\% as that obtained from the precision S2 orbit \citep{GRAVITY2018a}.  Alternatively, the two flares could also be {\bf on} the retrograde ISCO of a highly spun-up Kerr hole ($a\approx-1$, Figure \ref{fig:FigB3}, \citealt{kerr1963,bardeen1972}).

The \cite{bardeen1975}  effect is not expected to operate in a hot, geometrically thick accretion flow. So the face-on geometry we infer may be unrelated to the black hole spin direction, unless the accretion flow is brought into alignment for example by magnetic torques \citep{mckinney2012,sorathia2013}. \\
Our analysis and conclusions are fairly simple and empirically driven. The errors are substantial and the modelling by a single compact spot might be na\"{i}ve. If continuous energy injection dominates, the location of the hot spot may be affected by the propagation of that injection. It is certainly possible that in addition to rotation in a disk or torus, the hot spot may also have a line-of sight motion, perhaps due to a combined disk-jet geometry \citep{markoff2001,moscibrodzka2015}. However, the detection of orbital motion in bright flares is potentially evidence against radial motion in a jet being the dominant component (upper limit on motion perpendicular to orbit $<0.1c$; $<30\rm \,\mu as$); it also argues against Rossby wave instabilities with $m\ge2$ in the accretion flow \citep{falanga2007} as being the origin of the observed flux modulation. In summary, precision astrometry with GRAVITY at the VLTI {\bf provides strong support from spatially resolved measurements of the motions and polarization rotation of three strong near-IR flares in the relativistic accretion zone that SgrA* is indeed a massive black hole in the Schwarzschild-Kerr metric.}

\begin{acknowledgements}

We are very grateful to our funding agencies (MPG, ERC, CNRS, DFG, BMBF, Paris Observatory, Observatoire des Sciences de l'Univers de Grenoble, and the Funda\c{c}\~ao para a Ci\^encia e Tecnologia), to ESO and the ESO/Paranal staff, and to the many scientific and technical staff members in our institutions who helped to make NACO, SINFONI, and GRAVITY a reality. S.G., P.P. and C.D. acknowledge support from ERC starting grant No. 306311. F.E. and O.P. acknowledge support from ERC synergy grant No. 610058 (BlackHoleCam). J.D., M.B., and A.J.-R. were supported by a Sofja Kovalevskaja award from the Alexander von Humboldt foundation. A.A. \& P.G. acknowledge support from FCT-Portugal with reference UID/FIS/00099/2013.

\end{acknowledgements}

\bibliography{references}

\begin{appendix}

\section{}
\label{sec:AppA}
\subsection{Observations and analysis of GRAVITY data}
\label{sec:A1}

In this section, we summarise the May and July 2018 interferometric observations in the near-IR (2.2\,$\rm \mu m$) of SgrA* relevant to this paper.\\
The observations were taken at the VLTI in Chile using the recently deployed instrument GRAVITY \citep{GRAVITY2017}. The instrument coherently combines the light of the four 8m Unit Telescopes (UT) of the ESO Paranal site. We chose the most sensitive low resolution mode of GRAVITY. In this mode the science spectrum is dispersed across 14 pixels, with five independent spectral elements (R$\sim$20). All data sets were obtained in polarisation split mode, that is, a Wollaston prism is inserted in the optical train and the two linear polarisations are recorded independently. \\
Each observation followed the same sequence; all four UTs locked their Coud\'e IR adaptive optics (CIAO) module on the brightest source in the field, the red supergiant IRS7 ($m_K\approx6.5$, distance from SgrA* $\sim$5.5"). The interferometric observations started with IRS16NW feeding the fringe-tracker operating at a frame rate of 1kHz and IRS16C feeding the science channel with an integration time (DIT) of 1s. Those two bright stars ($m_K \approx 10.0 - 10.5$, separation from SgrA* $\sim$1") were used to find fringes and to zero the optical delay of the science channel. After this initial bright pair, we kept IRS16C as fringe-tracking star and only changed the science target (e.g. S2 and SgrA*) by moving the internal fibre actuators and rotating the field suitably. The science observations were interleaved exposures of NDIT=30 and DIT=10s each. We repeatedly pointed to the object R2, a moderately bright ($m_K \approx 12.1$, separation $\sim$1.5") nearby unresolved giant star, which served as a local calibrator and the S2/SgrA* binary. On average we took $\rm 5 \times 5$\,min SgrA* exposures before moving to the calibrator R2 and the science target S2/SgrA*. Every four or five exposures we interleaved a sky exposure. To do so, we offset the star separator field actuator located close to the Coud\'e focus by several arcseconds, pointing to a location devoid of stars in the Galactic centre.

\subsection{Data reduction}
\label{sec:A2}

We used the standard GRAVITY pipeline to process the data  \citep{2014SPIE.9146E..2DL,GRAVITY2017}. Each exposure consists of NDIT science frames, which are averaged after processing. Each individual frame is flat-fielded and corrected for a background bias by subtracting the closest sky exposure, detector noise, and wavelength calibrated on the internal calibration source. The data are then reduced based on a pixel-to-visibility matrix (P2VM, \citealt{2007A&A...464...29T}), which represents the matrix encoded instrument transfer function, including throughput, coherence, phase-shift, and cross-talk information for each individual pixel. In a second step the science complex visibilities are phase-referenced to the fringe-tracker complex visibilities using the laser metrology and accounting for the fibre dispersion to get phase-referenced visibilities. The observatory transfer function (i.e. coherence loss due to vibrations, uncorrected atmosphere, birefringence, etc.) was calibrated by observing a local calibrator (in our case the nearby giant star R2 at a distance of distance $\sim$1.5"). \\
We investigated the optimum NDIT number of science frames to average for the subsequent model fitting. We found that for this study, NDIT=3 frames (i.e. 30s) is a good compromise, which provides sufficient signal-to-noise ratio (S/N) and a reliable noise estimate, while at the same time allows the fast flux evolution of SgrA* to be resolved over timescales of minutes. Longer averages tend to smear out the signal due to rapid changes in brightness and/or separation.

\subsection{Model fitting}
\label{sec:A3}

The reported astrometric positions are based on a two-component binary fitting ($S_{\rm SgrA*}$ and $S_{\rm S2}$). We have developed several independent fitting codes, employing least-square minimisation with start parameter variation, Markov-Chain-Monte-Carlo (MCMC) optimisation, and a combination of both techniques. The goal was to have independent consistency checks, ensure robustness of the results based on various optimisation techniques, and to quickly explore new model parameters.
All codes are based on a binary model, which can be expressed in its most simple form by a complex visibility $V$ as

\begin{equation}
\label{eq:A1}
  V(u_k,v_k) = \frac{1+f_k e^{-2\pi i (u_k \cdot \Delta \alpha + v_k \cdot \Delta \delta)}}{1+f_k} 
,\end{equation}

where $f_k$ is the flux ratio of the two sources in the spectral channel, $\lambda_k$, $u_k$, and  $v_k$ are the spatial frequencies ($u_k={\rm u}/\lambda_k$  and $v_k={\rm v}/\lambda_k$, with u,v being the physical separation of the telescopes in east,north direction) and ($\Delta \alpha$, $\Delta \delta$) denotes the source separation vector (in right ascension and declination).
The red colour of SgrA* is usually expressed as a spectral slope $\nu F_\nu\sim \nu^\beta$ or alternatively $F_\lambda \sim \lambda^{-(1+\beta)}$ with typically $\beta \approx 0^{+0.6..-2}$  \citep{genzel2010,witzel2018,dodds-eden2011}. The slope of the early-type star S2 in the Rayleigh-Jeans limit is assumed to be $\nu F_\nu \sim \nu^3$ or $F_\lambda \sim \lambda^{-4}$.\\
Therefore, we account for the wavelength dependent flux ratio by

\begin{equation}
\label{eq:A2}
  f_k = F_{\rm SgrA*}/F_{\rm S2}\sim \left(\frac{\lambda_{k}}{2.2\, \rm \mu m} \right)^\gamma
  ,\end{equation}

where $\gamma$ relates to the intrinsic spectral slope $\beta$ of SgrA* by $\gamma=3-\beta$. \\
The short coherence length of the low-spectral-resolution mode and the comparably large separation of Sgr A* and S2 leads to a coherence loss, which is approximated with an optical delay $d$ dependent factor $\Gamma$ \citep{lachaume2013}

\begin{equation}
\label{eq:A3}
  \Gamma (d_k,R_k) = {\rm sinc} \left(\frac{d_k}{2R_k} \right)
  ,\end{equation}

with $d_k=2\pi \times ({\rm u} \Delta \alpha + {\rm v} \Delta \delta )/\lambda_k$ and the instrument resolution of a spectral channel $k$, $R_k = \lambda_k/ \Delta \lambda_k$. \\
The angular separation of S2 and SgrA* is comparable to the beam-diameter of the single-mode fibres ($\sim$50 mas). Therefore, we have to consider the relative injection of S2 and SgrA* per telescope (alignment errors can lead to a different injection ratio). Based on auxiliary data we can infer the telescope $t$ dependent injection $I_{t,\rm SgrA*}$  and  $I_{t,\rm S2}$ based on the fibre- and object separation. We write the injected flux ratio of telescope $t$ and spectral channel $k$ as 

\begin{equation}
\label{eq:A4}
f_{t,k}=\frac{I_{t,\rm SgrA*}}{I_{t,\rm S2}} f_k
  ,\end{equation}

with the telescope $t$ and the intrinsic (wavelength dependent) flux ratio $f_k=f_{2.2} (\lambda_k / 2.2\mu m)^\gamma$ and obtain, in the most basic model, the complex visibility for telescope 1,2 for one spectral channel $k$,

\begin{equation}
\label{eq:A5}
  V(u_k,v_k) = \frac{1+ \Gamma(d_k,R_k) \sqrt{f_{1,k} f_{2,k}} f_k e^{-2\pi i /\lambda_k ( {\rm u} \cdot \Delta \alpha + {\rm v} \cdot \Delta \delta)}}{\sqrt{1+f_{1,k}} \sqrt{1+f_{2,k}} } 
.\end{equation}

\begin{figure}
\includegraphics[width=9cm]{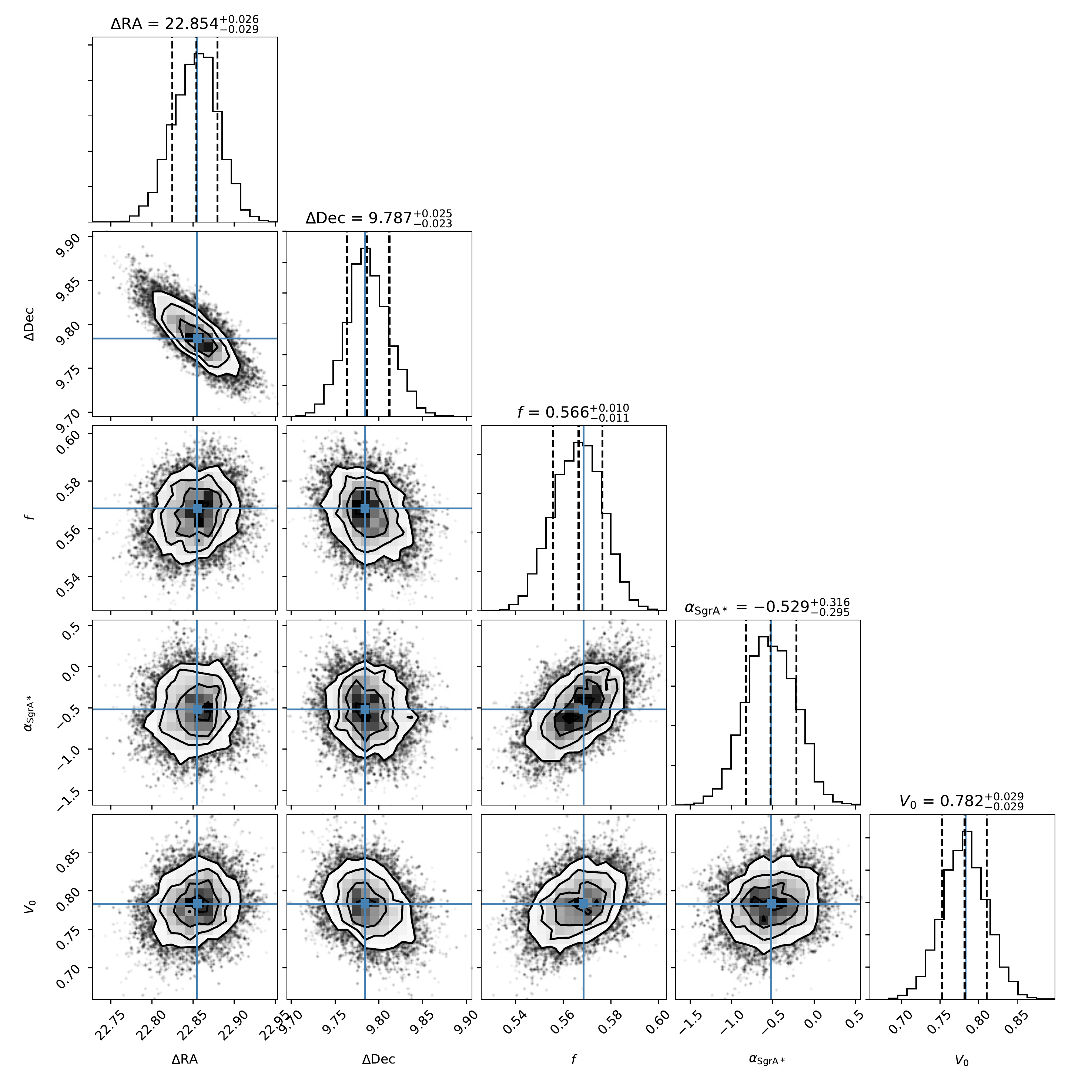}
\caption{Example corner plot from the `Waisberg' MCMC analysis showing the marginalised likelihood distributions over the five parameters of the fit for one single 30s frame during the July 22 flare: $\Delta$RA and $\Delta$Dec positions, Sgr A* / S2 flux ratio $f$ in the photometric fibre, the Sgr A* spectral index $\alpha$, and the unresolved background level $V_0$ in the data.}
\label{fig:FigA0}
\end{figure}

Some of the four analysis codes integrate over the bandpass and allow for a colour-dependant background. Overall the results are in excellent agreement and are independent of the optimisation technique and the detailed implementation. Figure \ref{fig:FigA0} shows one example MCMC fit of the July $\rm 22^{}$ `Waisberg' analysis. 

\subsection{Linear polarization analysis}
\label{sec:A4}

During all observations, a Wollaston prism was introduced into the optical beam between the integrated optics chip and the dispersive grism and the detector \citep{GRAVITY2017}. This setup splits the light into two orthogonal polarization components, which are imaged onto the detector. The Wollaston prism is aligned with the axes of the integrated optics chip to avoid polarization crosstalk from the intrinsically birefringent chip. At the entrance of the GRAVITY instrument, a half-wave plate is located in the beam. During observations the half-wave plate co-rotates with the field de-rotator device. This ensures that the incident angle of the internal laser metrology is fixed with respect to the derotator mirrors and thus avoids birefringence-related phase-shifts and corresponding astrometric biases. This however leads to a rotation of the detected polarization components with respect to the sky N-E coordinate system. Therefore the measured polarization angle needs to be corrected for the field rotation at the instrument.  \\
In order to analyse the linear polarization properties of a source on sky with an arbitrary position angle, it is common practice to split the polarization and to probe the polarization in at least two rotated states. In our case we offset the co-rotating half-wave plate by $0^{\circ}$ and $45^{\circ}$ during science and calibrator exposures. Each $0^{\circ}$ and $45^{\circ}$ science exposure is calibrated with the corresponding calibrator exposures.\\
A single exposure provides two orthogonal polarization states. Consequently, the half-wave plate at $0^{\circ}$ yields the Stokes parameter $Q=P_0 - P_{90}$  and the $45^{\circ}$ state provides the parameter $U=P_{45}-P_{135}$. The total intensity is $I=P_{0}+P_{90}=P_{45}+P_{135}$. The degree of polarization can be inferred from $\Pi=\sqrt{U^2+Q^2 }/I$ and the polarization angle is defined as $\theta=1/2 {\rm arctan}(U/Q)$. \\
We tested and calibrated the linear polarization measurement with GRAVITY on sources with known polarization properties (taken from \citealt{ott1999,buchholz2013}), namely GCIRS21 ($P\approx14\%$; $\theta\approx15^{\circ}$), IRS16SW ($P\approx3.1\%$; $\theta\approx20^{\circ}$), GCIRS33E ($P\approx5.7\%$; $\theta\approx35^{\circ}$) and IRS1W ($P\approx1.8\%$; $\theta \approx-37^{\circ}$). Here the polarization angle is defined in the range $[-90^{\circ},90^{\circ}]$ with the angle increasing east of north. Based on comparison with the aforementioned reference stars, we estimate the systematic uncertainty of the polarization degree measurement to be $\Delta P\approx1.5\%$ and the angular uncertainty to be $\Delta \theta \approx10^{\circ}$. 
It should be noted that the polarization measurement of SgrA* is actually a differential measurement, since it is based on the flux ratio between S2 and SgrA*. Instrumental polarization effects, which are common to both sources should cancel to first order. \\
One caveat of measuring the polarization angle and degree with the technique of rotating a half-wave plate is that the two measurements are done sequentially. This requires interpolating Q and U over the full observing sequence to fill the respective gaps. A typical exposure takes 5 min per state and the overheads to rotate the half-wave plate take another $\sim$1\,min. This means that the gaps are roughly 6 min. The polarization measurement is effectively smoothed over short timescales. On timescales of a flare ($\ge$30\,min), the time resolution is sufficient to capture the evolution of the polarization.

\subsection{Field rotation}
\label{sec:A5}

\begin{figure}
\includegraphics[width=9cm]{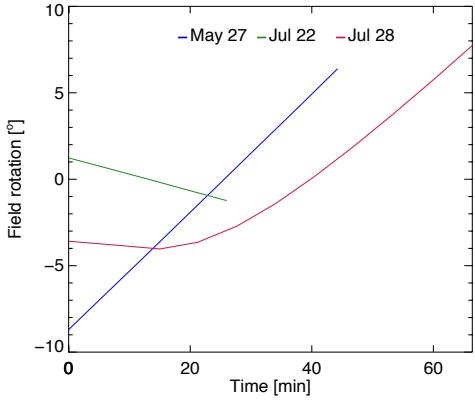}
\caption{Field rotation (mean angle subtracted) at the instrument during the three flares.}
\label{fig:FigA1}
\end{figure}

One potential source of systematic error is the field rotation during an observation. In particular polarization can be affected due to changing angles of incidence as the mirrors track the source on sky. In principle this is calibrated by repeated exposures on a calibrator star, yet some signature might still be present in the data. Figure \ref{fig:FigA1} shows the evolution of the field angle at the instrument (subtracted by the mean angle) during the three flares. The field angle at the instrument changes only slowly during all three flares. In particular, during the flare with polarization information (July 28) the angle changed by less than $14^{\circ}$, and on July 22 by only $3^{\circ}$. On the short timescales of the flares the field rotation is unlikely to impact the polarization and astrometry at a significant level.

\subsection{Astrometric precision and accuracy}
\label{sec:A6}

\begin{figure}
\includegraphics[width=9cm]{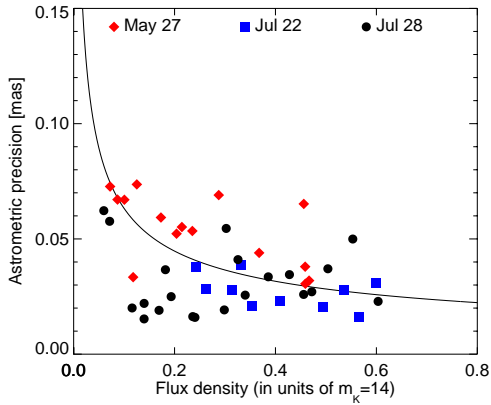}
\caption{Astrometric precision as a function of the flux density of SgrA*. Each point represents the 2D standard deviation of six individual 30-second position fits within 3 minutes, as in Figs. 1 and 2, divided by $\sqrt{6}$ . Here both polarizations were averaged. The contiguous line indicates a photon-noise-limited phase noise \citep[e.g.][]{wyant1975} $\sigma_{\phi}\propto1/\sqrt{N_{ph}}$, where $N_{ph}$ is the number of photons.}
\label{fig:FigA2}
\end{figure}

We derived the astrometric precision based on the scatter of the individual fit results. We fitted individual 30s exposures (3 frames with 10s DIT). The results were binned over 3 minutes (i.e. $\sim$6 positions) resulting in the final astrometric positions presented in Figs. \ref{fig:Fig1} and \ref{fig:Fig2}. Consequently we estimated the astrometric precision from the scatter within each bin. Figure \ref{fig:FigA2} shows the precision as a function of SgrA* brightness. For the brightest states, we achieve a 1D precision of $\rm 20\,\mu as$ and a 2D precision of $\rm \sim30\,\mu as$ rms.\\
The long-term stability of the wavelength calibration of GRAVITY in low spectral resolution has been measured to be better than $0.45$\,nm rms (over 6 months). This corresponds to $2.0\times 10^{-4}$ in relative terms; a negligible contributor to the uncertainties.

\subsection{Systematic astrometric error sources}
\label{sec:A7}

\subsubsection{Baseline uncertainty}

From Equation \ref{eq:A1} it is obvious that the measured separation $\Delta \alpha$, $\Delta \delta$ depends on the precise knowledge of the baseline vector (u,v). Any uncertainty in the baseline ($\delta \vec{B}$) directly translates into a separation error $\delta \vec{S}$ as

\begin{equation}
\label{eq:A6}
\frac{\delta \vec{B}}{\vec{B}}=\frac{\delta \vec{S}}{\vec{\Delta S}}
.\end{equation}

The knowledge of the baseline vector is limited by several terms such as the global uncertainty of the telescope array geometry and the knowledge on the actual pointing vector to the object (i.e. an uncertainty in R.A. and Dec.). The global array geometry only refers to the separation of the pivot points of the telescopes. Unlike in radio interferometry, where the separation of the telescopes is much larger than the size of the telescopes and the telescope can be considered point-like in the baseline u-v space, in optical interferometry the telescopes size can play a significant role. This means that optical errors such as pupil mis-registration or pupil vignetting can contribute to baseline errors. \\
The typical optical path accuracy of the VLTI is $\rm <1\,cm$ (i.e. the fringes are found within $\rm <1\,cm$ from the predicted position) during a preset to objects, where accurate positions are available. This means that the global array geometry has to be accurate to the same level. The absolute positions of SgrA* is extremely well determined based on a radio reference frame, which is tied to nearby Quasars (accuracy $\rm <1\,mas$, \citealt{reid2014,plewa2015}). This leads to a negligible baseline error of $\rm 1\,mas \times 100\,m \approx 10^{-4}\,cm$.
In order to minimize the pupil error, the GRAVITY instrument is equipped with a dedicated laser guiding system, which controls pupil runout. The guiding system centres the pupil with a residual 1D scatter over 5 minutes of $\rm \sim0.5\,cm$ rms. Constant monitoring of the pupil illumination argues against baseline errors due to vignetting. Overall the baseline error is at maximum on the order of $\rm \sim1\,cm$. For a 100\,m baseline the corresponding systematic astrometric error for two sources separated by $\rm{\sim}20\,mas$ is $\rm \delta \vec{S}\approx20\,mas \cdot10^{-2}/100\approx 2\mu as$. The small separation between SgrA* and S2 in 2018 leads to a negligible contribution of baseline errors in the relative astrometry.

\subsubsection{Effective wavelength accuracy}

Another systematic uncertainty is related to the effective wavelength (see Equation \ref{eq:A1}). The wavelength determines the image scale for an interferometer. This means that any wavelength uncertainty leads to a proportional astrometric uncertainty. In the low-resolution mode used for the observations the instrument features 14 spectral channels across the K-band ($\rm 2.0-2.45\,\mu m$). The effective bandwidth of each channel varies from FWHM $\rm{\sim}70 ~to~ 140\,nm$, with significant overlap between the channels. The large bandwidth of the spectral channels can lead to a shift of the effective wavelength for objects, which are significantly redder than the relatively blue calibration lamp.
The average H-K colour in the Galactic centre is 1.8mag \citep{fritz2011}. This steep colour slope from extinction can be approximated by a power-law dependence of $E(\lambda)_{2.2}\propto \lambda^{-5}$ in the K-band. The strong extinction leads to an effective colour temperature of a B0V type star such as S2 of only $\rm {\sim}1000\,K$. In comparison the calibration lamp with a colour temperature of ${\sim}2800\,K$ appears blue. Considering the exact bandpass shape, we calculated an overall effective wavelength shift of $\rm {\sim}2\,nm$,that is, an image scale change of $\rm \approx 2\,nm/2200\,nm =0.1\%$. This overall shift is taken into account for the final astrometry.  However, the spectral slope of SgrA* is intrinsically redder than S2 and potentially varies with the state of SgrA* between $\beta=-2...+0.6$ \citep{genzel2010}. This leads to an additional effective wavelength uncertainty, which is of the order $\rm {\sim}1\,nm$, that is, 0.05\%. The corresponding astrometric uncertainty is $\rm {\sim}10\,\mu arcsec$, assuming 20\,mas separation between S2 and SgrA*.

\subsubsection{Atmospheric limitations}

Due to a small non-common atmospheric path, the two objects are subject to differential tip-tilt. The order of magnitude can be approximated from \citep{shao92},

\begin{equation}
\label{eq:A7}
\sigma_{tt}=540 B^{-2/3}  \Theta \cdot t^{-1/2}~~\rm [arcsec]
\end{equation}

with $B=100\,\rm m$, object separation $\Theta=1 \cdot 10^{-7}\, \rm rad$ (20\,mas) and $t=100s$, the residual tip-tilt is insignificant at the level $\sigma_{tt}\approx0.2\, \rm \mu arcsec$.

\subsubsection{Atmospheric refraction}

Atmospheric refraction leads to a wavelength-dependent shift of the observed zenith angle relative to the true zenith angle. For a single telescope this shift needs to be considered for an accurate pointing and astrometry.
The atmospheric refraction is defined as $R\coloneqq z_t-z_a$ with the true and apparent zenith distance $z_t$ and $z_a$. The wavelength-dependant refraction causes astronomical objects to appear dispersed into a spectrum along the parallactic angle. The refractive index of air at $\rm 2.2\,\mu m$ for typical Paranal conditions (pressure=744\,hPa, temperature=$\rm 10^{\circ}C$, humidity=10\%) is $(n_{2.2}-1)\times10^6=203.95$ and the dispersion is $dn/d\lambda |_{2.2}= -2.21\cdot10^{-10}  ~\rm [nm^{-1}]$. The corresponding differential refraction across the K-band ($\rm2.0-2.45\,\mu m$) leads to an angular dispersion of $\Delta R\approx 20\,{\rm[mas]}\, {\rm tan}(z_t)$. This can lead to a loss of injection into the single-mode fibres for large zenith angles.\\
For interferometers such as the VLTI the atmosphere can be considered uniform and plane parallel (earth curvature can be neglected, e.g. \citealt{mathar2005}).
The refraction as the rays travel through the atmosphere is governed by Snell's law, that is, $n_0\,  {\rm sin}(z_0)=n\, {\rm sin}(z) $ is constant.
The corresponding delay measured in an interferometer $D=b \,n_0\,  {\rm sin}(z_0)=b\,n\, {\rm sin}(z)$ is constant and is equal to the delay in free space \citep{thomson2017}. In the case of an interferometer with evacuated delay lines, atmospheric refraction only needs to be considered for the pointing of the telescopes but has no effect on the interferometric measurement.\\

The delay lines of the VLTI are however in air and therefore dispersion is introduced due to the optical path compensation. The effect of dispersion in the delay lines is exactly the same as differential refraction on sky.\\
The colour difference between S2 and SgrA* leads to a differential refraction between the two objects for observations at non-zero zenith angles.
Across the K-band the differential refraction in the direction of the parallactic angle (with stronger refraction in the blue) is

\begin{equation}
\label{eq:A8}
\Delta R= 45 \,({\rm \mu as ~nm^{-1}}) \cdot \Delta \lambda ~ {\rm tan}(z_t)
,\end{equation}

for a wavelength shift of $\Delta \lambda$. The three flares discussed here were all observed at zenith angles $z_t<35^{\circ}$ and changed during the observation by less than $10^{\circ}$. We assume an average colour for SgrA* of $\beta=0$, which corresponds to an effective wavelength shift of 1\,nm relative to S2 for 100\,nm the bandwidth of the instrument resolution. We take the corresponding astrometric shift into account. The relative shift during the flares due to changing zenith angle is in all cases smaller than $\rm 15\,\mu arcsec$.  Relevant for the flare astrometry is only a change in differential refraction. The effective wavelength uncertainty of SgrA* of up to $\rm {\sim}1nm$ (due to its varying spectral slope), translates into a maximum differential refraction variation of $\rm {\sim}26\,\mu as$.

 \section{} 
\label{sec:AppB}
\subsection{Data Modelling}
\label{subsec:data_model}
\subsubsection{Model limitations}
\label{subsec:model_lim}
\begin{figure*}[ht]
\includegraphics[width=18cm]{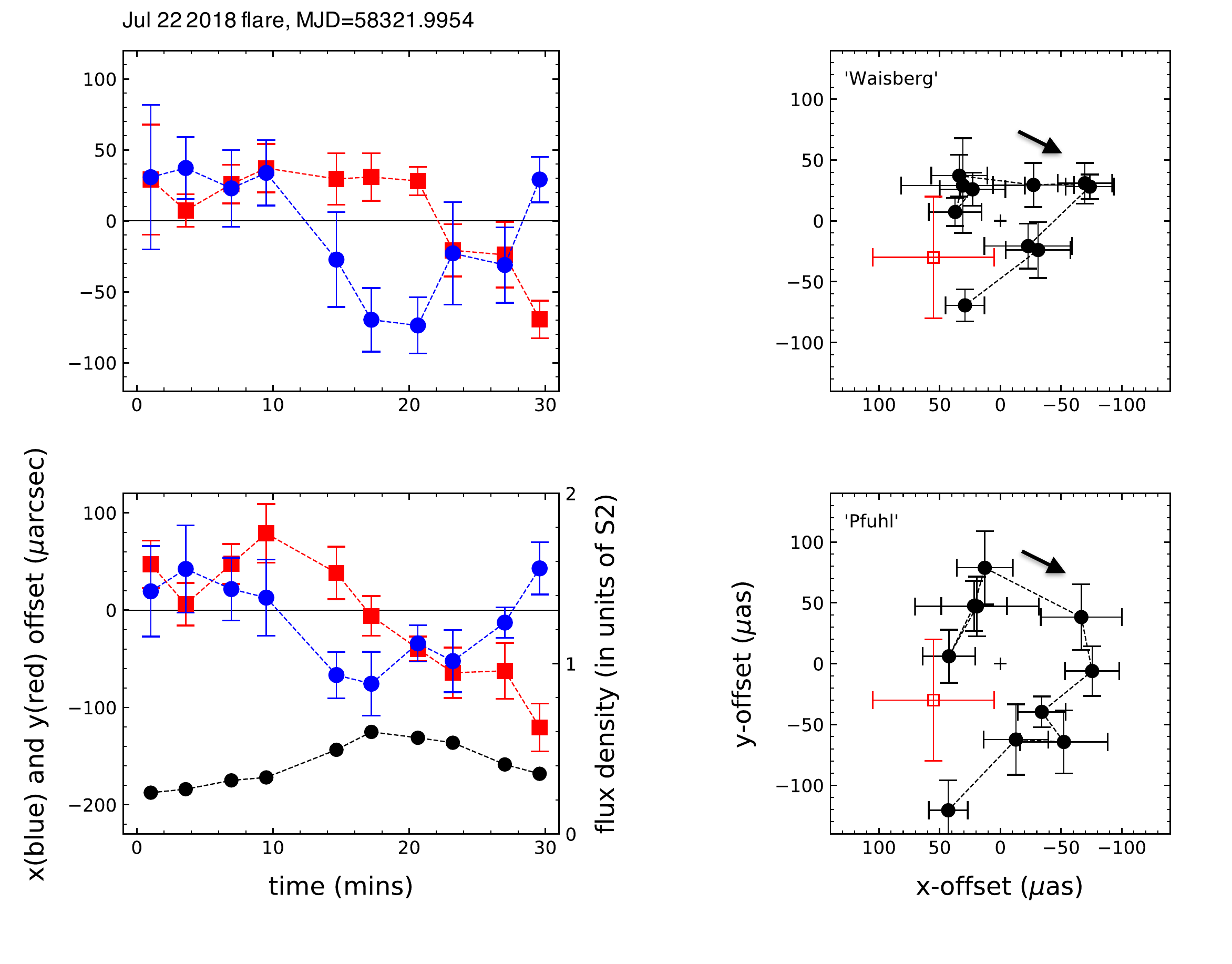}
\caption{Comparison of the Pfuhl (bottom) and Waisberg (top) analyses of the July 22 flare data. Nomenclature as in Figure \ref{fig:Fig1}.}
\label{fig:FigB1}
\end{figure*}

We developed four independent codes to fit the coherent signal of SgrA* and S2 and at the same time account for instrumental properties. The codes differ in the fitting approach (MCMC, least-square with parameter grids, etc.), the number of free parameters (e.g. intrinsic source spectral slope is fixed or fitted), and the weighting of closure phase to visibility data. The different choices in the codes are all well motivated and can thus be considered as an exploration of the parameter space. Figure \ref{fig:FigB1} shows the results of two codes named "Waisberg" and "Pfuhl" for the $\rm July \,22^{}$ flare (Figure \ref{fig:Fig1}). The "LESIA" results are nearly identical to the Waisberg results. The "Horrobin" code shows differences on a similar scale to Pfuhl and Waisberg. While the four codes agree on the main results and features, they show also differences, which reflect another source of systematic uncertainty on the order $\rm 20-30\,\mu arcsec$.

\subsection{Astrometric modelling of orbits}

\begin{figure*}[ht]
\includegraphics[width=18cm]{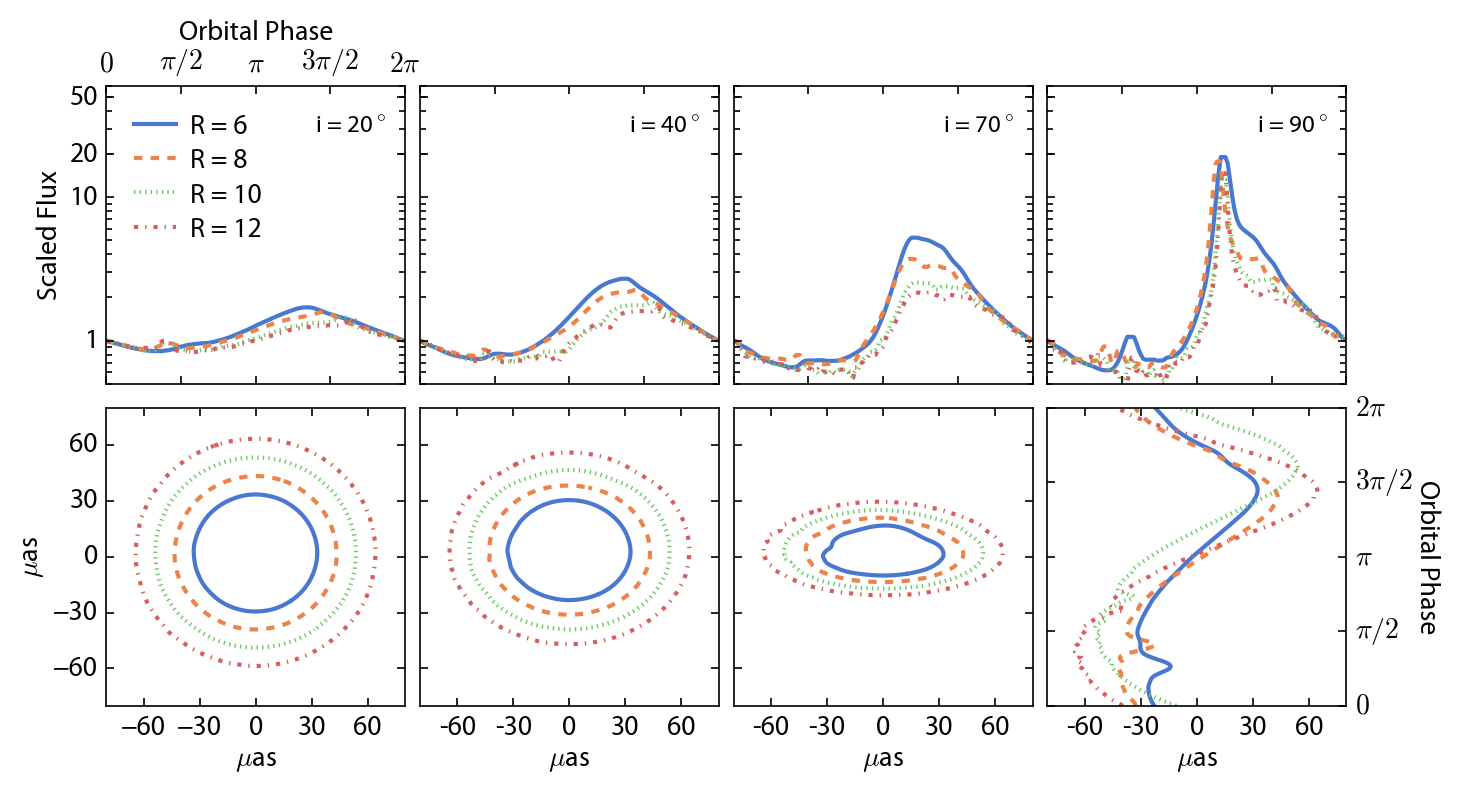}
\caption{Light curves (top) and centroid tracks (bottom) of a compact hot spot orbiting a Schwarzschild black hole for four different orbital radii and inclinations ($R/R_g$=6 (solid blue), 8 (red dashed), 10 (short-dashed green) and 12 (dash-dotted red)). From left to right the inclinations are 20, 40, 70 and $90^{\circ}$. Since the ordinary centroid for the $90^{\circ}$ case simply yields a horizontal line, we instead plot the orbital phase $\Phi(t)$ against $x(t)$ (based on the { \bf NERO} code (Baub\"ock et al. in prep.); see also \citealt{hamaus2009}).}
\label{fig:FigB2}
\end{figure*}

Figure \ref{fig:FigB2} shows models of hot spots on circular orbits in the Schwarzschild metric (angular momentum parameter $a=0$) obtained with ray tracing methods of the geodesics, including secondary, tertiary, and quaternary images (\citealt{hamaus2009}, see also \citealt{broderick2005,broderick2006,vincent2011}) . The same reference also shows models for the Kerr metric. Similar models with or without polarization have been published elsewhere \citep{broderick2005,broderick2006,hamaus2009,vincent2011,vincent2014}. The $a=0$ orbits for inclinations $<20^{\circ}$ near ISCO and at somewhat higher inclinations for R$>$ISCO result in spot centroid motions that are little influenced by light bending and multiple images. The reverse is true for high spin, highly inclined orbits. Given the smooth near-circular spot motions at least for the July $\rm 22^{}$ data (Figure \ref{fig:Fig1}), the data are consistent with the low inclination orbits in Figure \ref{fig:FigB2}.\\
To simulate the effects of astrometry noise on these simple near-circular orbits, we added a randomly drawn $\delta x$ and $\delta y$ noise component to each of the model data points, with  a magnitude comparable to the average empirically determined noise in each of the flares $\rm 20-25\,\mu as$ 1D for July $\rm 22^{}$, and $\rm 45\,\mu as$ for July $\rm 28^{}$ and May $\rm 27^{}$. We assumed that the orbits obey the radius R-period P relations of circular orbits in the Schwarzschild-Kerr metric,

\begin{equation}
\label{eq:A9}
P=2.137 {\rm (min)} \times \left(\frac{M_{\bullet}}{4.14\times10^{6} M_{sun}}\right) \times  \left(a + \left({\frac{R}{R_g}}\right)^{3/2}\right)
,\end{equation}

where $a$ is the spin parameter $ \in [-1,1]$, and $R_g=6.11\times10^{11} \,[M_{\bullet}/4.14\times10^6 M_{\odot}] ~{\rm cm}=5\rm \,\mu as$. We then varied $R$ to obtain a reasonable match with the data (and their uncertainties) in the three flares. As we discuss in the main text, we find that an orbit on a common radius, $R\approx1.17 \pm0.25 \,R_{\rm ISCO}(a=0)$ can match all three flares (Figure \ref{fig:Fig3}). However, due to the fact that only a partial orbit is observed, we cannot rule out non-circular orbits. \\
We model the astrometry data with fully relativistic models of a compact distribution of gas orbiting at a common velocity in the Kerr spacetime (a "hotspot", \citealt{hamaus2009,vincent2011,vincent2014,connors1980}). When the orbit is in the equatorial plane of the black hole, the motion can be calculated analytically with angular frequency $d\phi/dt = \Omega = (r^{3/2}+a)^{-1}$ in Boyer-Lindquist (BL) coordinates. 
Ray tracing is used to account for relativistic effects. From an observer's camera photon geodesics are traced backwards towards the black hole. The hotspot density is assumed to fall off as a Gaussian with distance from its centre, with a characteristic size $R_{\rm spot} << R$. The intensity along the ray is then calculated assuming the emission is optically thin and taking into account the frequency shifts due to Doppler beaming and gravitational redshift.\\
We have used three different codes for this technique. All codes allowed for optimisation of the orbit radius, inclination, position angle, and phase angle of the spot at $t = 0$. The spin parameter was fixed to either 0 or -1, corresponding to counter-rotating around a maximally spinning black hole. In {\bf grtrans}\footnote{{\bf grtrans} is available from: $\rm https://github.com/jadexter/grtrans$}  \citep{dexter2016,dexter2009}, the emissivity is polarized synchrotron radiation from a power-law distribution of non-thermal electrons \citep{broderick2006}. The ray tracing technique accounts both for the emitted polarization and its parallel transport to the observer. \\
In {\bf GYOTO}\footnote{{\bf GYOTO} from: $\rm https://gyoto.obspm.fr$}  \citep{vincent2011,vincent2011b,grould2016}, the source is an optically thick coordinate-sphere of radius 0.5 (in BL coordinates) with emissivity $I_\nu(\nu) =$ constant.  In this set-up, the power-law index would only matter for weighing the contribution of the secondary image, which is minor at such low inclination. The source is set on a circular orbit in the equatorial plane of a Kerr black-hole and we ray-trace the appearance of the sphere at the actual observing dates over a field of view of 200\, $\rm \mu$as with a pixel size of 1\,$\rm \mu$as. The model astrometry is then the centroid of these ray-traced images. We minimise the free parameters with the standard Python procedure {$scipy.optimize.curve\_fit$}. The distance of the GC and mass of the central object are fixed at $M_{\rm BH}=4.14\times 10^6 M_{\odot}$ and $R_0=8.127$ kpc \citep{GRAVITY2018a}.  \\
In the {\bf NERO} code (Baub\"ock et al. in prep.), we combine the {\bf YNOGKM}\footnote{{\bf YNOGKM} from: $\rm  {http://www1.ynao.ac.cn/{\sim}yangxl/yxl.html}$} \citep{yang2014} code to calculate the timelike geodesics of particle orbits with the {\bf geokerr}\footnote{{\bf geokerr} from: \\$\rm https://faculty.washington.edu/agol/geokerr/index.html$} \citep{dexter2009} code to find the null geodesics of photon trajectories. Here we consider only models of a small hotspot with a Gaussian density profile on a circular orbit around a non-spinning ($R\ge 6\, R_g$) or maximally spinning ($R<6\, R_g$) black hole. We image the resulting orbit by ray-tracing photon trajectories in a field of view corresponding to 1.5 times the size of the orbital radius. We calculate the flux at each point by integrating the density along the photon path, that is, the emissivity is proportional to the density. By accounting for the time of travel of the photons as well as the hotspot motion, we can calculate a time-dependent image of the region near the black hole. The position-weighted average of the flux of this image gives the centroid of the emission.\\
We fit the resulting centroid tracks to the data over a parameter space that spans the radius and inclination of the orbits, the position angle of the angular momentum vector, and the phase of the point at $t=0$. We employ a grid fit of models between $R=3 \, R_g$ and $R =12\, R_g$, where those orbits that fall outside the ISCO are modelled with a zero-spin black hole, while at $R < 6\, R_g$ we use prograde orbits around a maximally spinning black hole. \\
All codes find the same relativistic effects. Lensing causes the orbit to appear somewhat larger on the sky. At higher inclination, strong Doppler beaming causes large flux modulations between the approaching and receding sides of the orbit with an additional peak from lensing behind the black hole. The spot appears to move faster when approaching than when receding. 

\begin{figure}
\includegraphics[width=8.5cm]{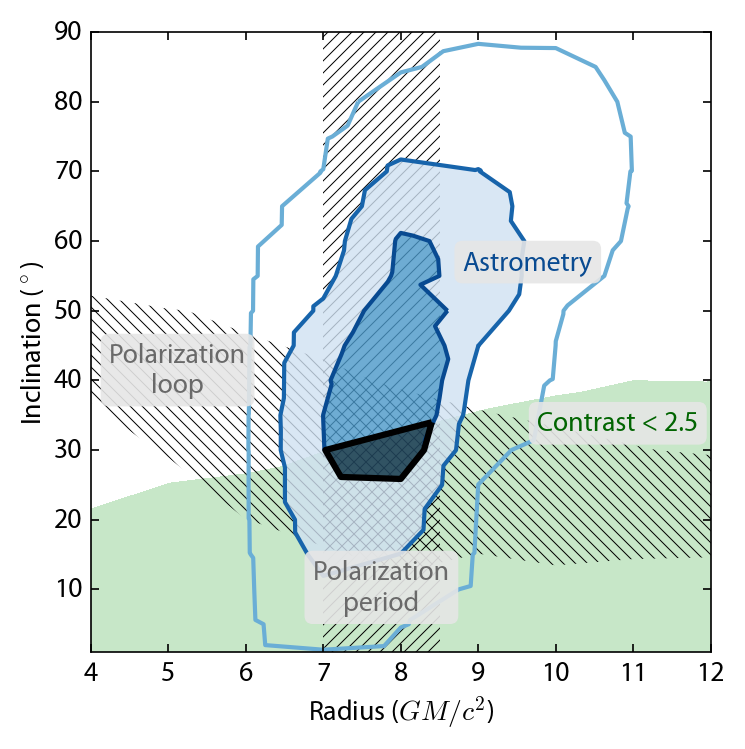}
\caption{Combined constraints of the location (astrometry) of the July $\rm 22^{}$ flare in the $R/R_g$ - inclination plane, from fits with the {\bf NERO} code (Baub\"ock et al. in prep. but also \citealt{yang2014,dexter2009}). For $R\ge6\, R_g$ we use $a=0$, and for $R < 6\, R_g$, $a=1$. The three astrometry contours represent the 1-, 2- and 3-$\sigma$ constraints from the model fitting, with the dark blue being the most favoured. Since we do not have any prior knowledge of the inclination of the orbit, we have weighted the probabilities by $\rm {\rm sin}(i)$ to account for the geometric bias. The polarization measurement of July 28$\rm ^{}$ with a period of $P_{\rm pol}=48\pm6$ min yields the right-hatched vertical constraint. The requirement of a Doppler boost (`contrast') $<2.5$  is given by the green area, with a lower contrast pushing further down. The presence of a single and full loop in the polarization data yields the left-hatched, mostly left-right constraint. The combined constraints favour $R=7.6\pm0.5\, R_g$ and inclination $i\le30$ degrees (black encircled area).}
\label{fig:FigB3}
\end{figure}

\subsection{Hot spot models of the May 27 and July 28 flares}
\label{subsec:may27}
\begin{figure*}[ht]
\includegraphics[width=18cm]{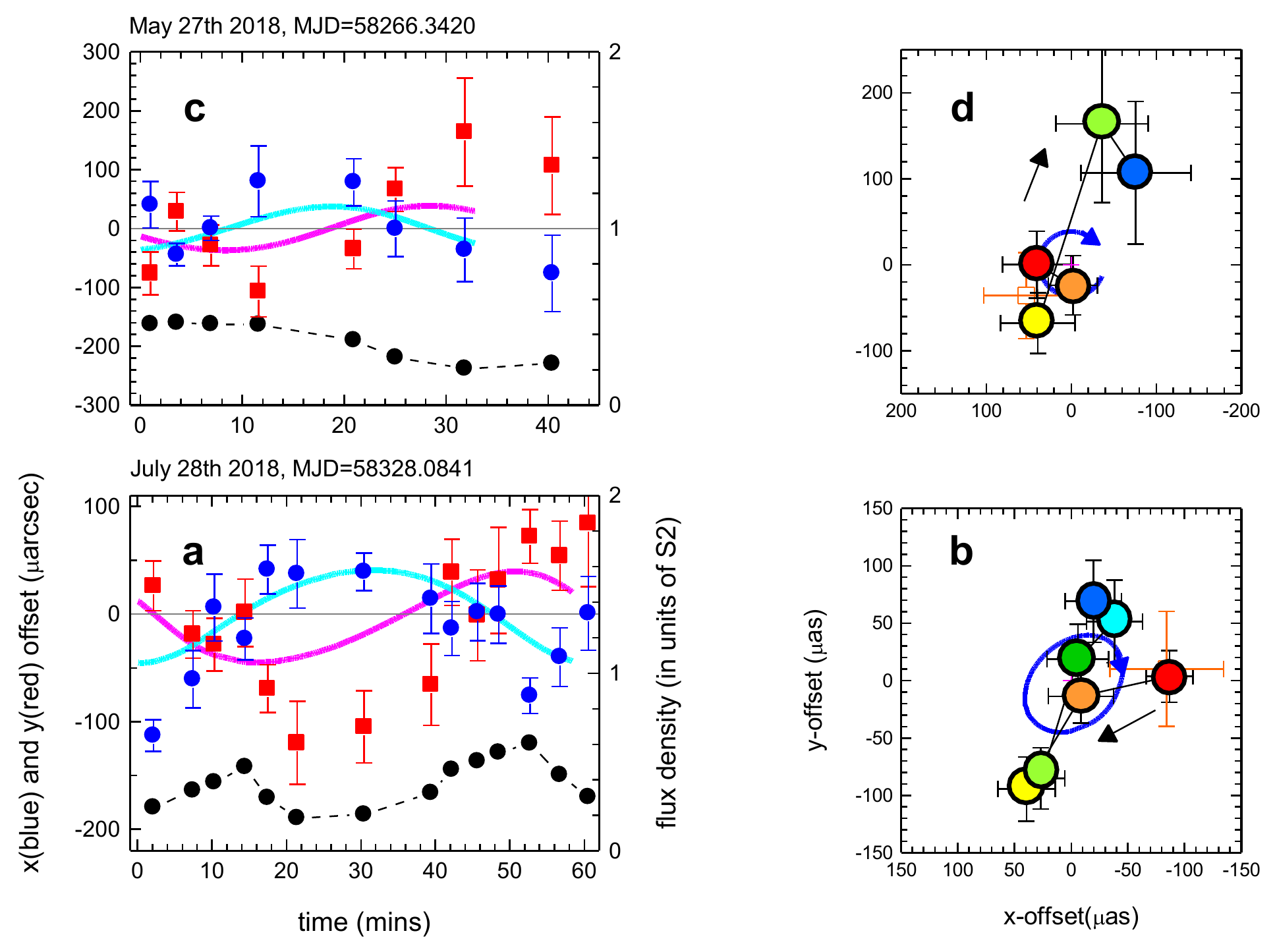}
\caption{Bottom left (a): time evolution of the east-west (east positive, blue) and north-south (red) position offset of the July 28 (MJD=58328.0841) centroids from their medians, as well the flux density evolution (right y-axis, black), in units of the flux of S2 (14.0 mag). Error bars are 1$\sigma$. For this purpose the total intensity was computed from the sum of the two polarization directions. The data points use the  `Pfuhl' analysis (see Appendix B). Bottom right (b): Projected orbit of the flare centroid on the sky with colours from red to blue marking the time evolution. The orange square indicates the black hole position including long-term astrometric uncertainty.  The thin continuous blue curve denotes a simple model of a hot spot on a circular orbit with $R=9 \,R_g$ ($a=0$, $M=4.14\times 10^6 M_{\odot}$), seen at inclination $135^{\circ}$ (clockwise on the sky, as for the data (black curved arrow)) and with the line of nodes at $\Omega=120^{\circ}$ (same as in Figure \ref{fig:Fig1}), fitted with {\bf NERO} code (Baub\"ock et al. in prep. and \citealt{yang2014,dexter2009}). The cyan and violet curves show the orbit in $x(t)$ and $y(t)$. Top (c and d): As in (a) and (b) but for the May 27 flare (MJD 58266.3420). Here the best fitting orbit is face on with $R=7\, R_g$.}
\label{fig:FigB4}
\end{figure*}

We have analysed the centroid spot motions of the May 27 and July 28 flares with the same techniques as for the July 22 flare. The results are given in Figure \ref{fig:FigB4}.

 \section{Limit on inclination of orbits based on the lack of Doppler boosting signal} 
\label{subsec:doppler}
We have discussed in the main text the lack of any obvious pronounced brightening of the flux evolution in one particular section of the three orbits. If a relativistic hot spot is moving with $\beta=v/c$ and $ \Gamma=(1-\beta^2)^{-1/2} $ at an angle $90-i$ relative to the line of sight, emitting a spectrum with spectral index $\alpha$, $S_{\nu}\propto\nu^{-\alpha}$, then the combination of relativistic aberration, time dilation, and Doppler frequency shifting boosts or de-boosts the observed flux density in the stationary observer's frame by a factor \citep{mccrea1972}:

\begin{equation}
\label{eq:A10}
S_{\nu,{\rm obs}}=\nu^{-\alpha} \times \left[\Gamma \times (1-\beta \, {\rm cos}(90^{\circ}-i) ) \right]^{-(3+\alpha)}
.\end{equation}

\begin{figure}
\includegraphics[width=9cm]{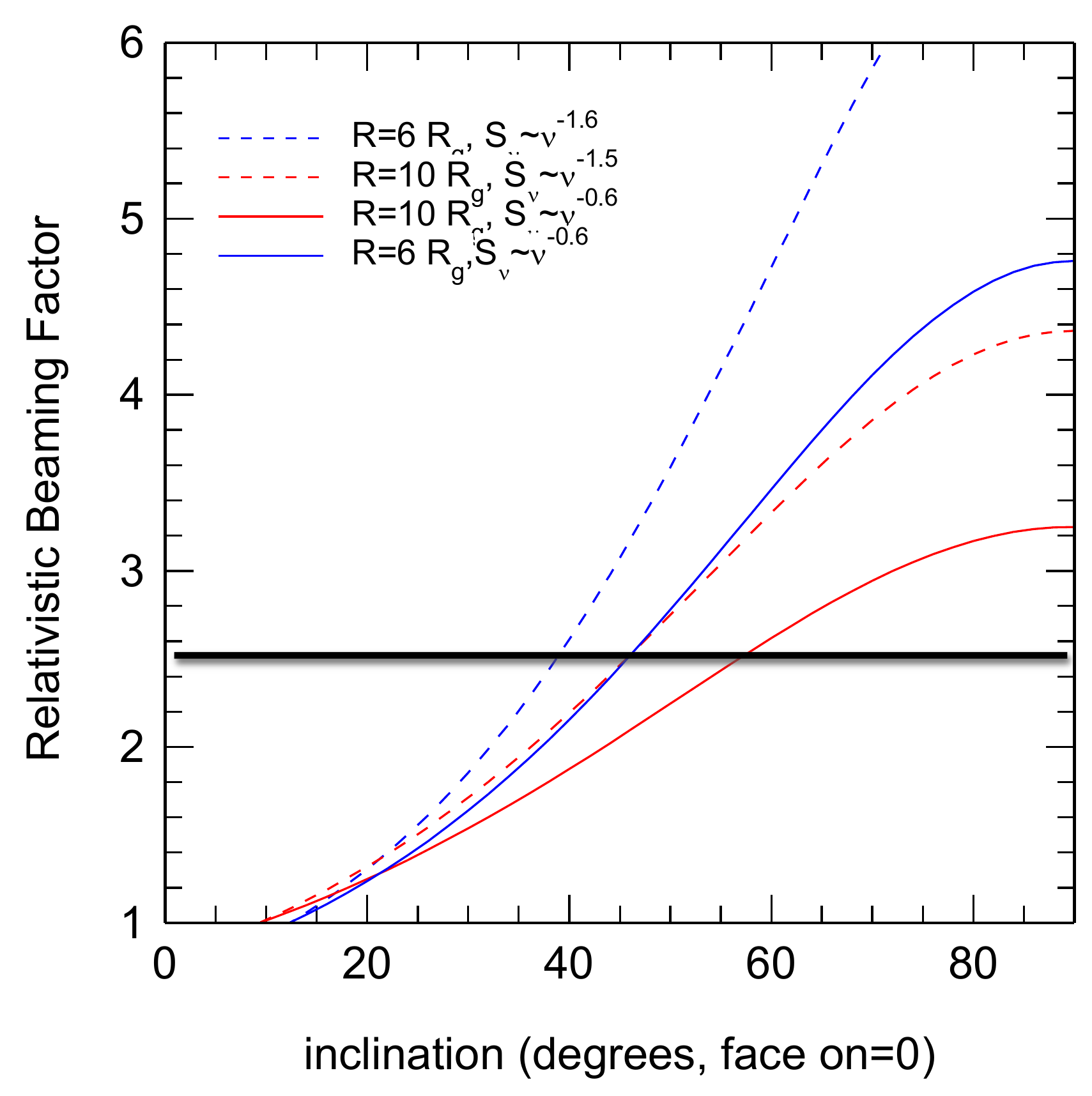}
\caption{Doppler boosting factor in the observer's frame $o$ relative to the hot spot's moving frame $hs$, $S_o \sim S_{hs}\times D$, with $D=(\Gamma(1-\beta))^{-(3+\alpha)}$, for a moving hot spot at $\beta=v/c$, $\Gamma=(1-\beta^2)^{-1/2}$ , and at an angle $90^{\circ}-i$ relative to the line of sight \citep{mccrea1972}. We show $D$ for two radii, ISCO $a=0$ (blue) and $R=10\, R_g$ (red). Solid curves assume spectral index $\alpha=0.6$ ($S_{\nu}\propto\nu^{-\alpha}$, appropriate for very bright flares) and dotted curves assume $\alpha=1.5$ ($S_{\nu}\propto\nu^{-\alpha}$, appropriate for average bright states, \citealt{witzel2018}). If one assumes that $D<2.5$, the inclination has to be less than about $45^{\circ}$.}
\label{fig:FigC1}
\end{figure}

Figure \ref{fig:FigC1} shows these `boosting' factors as a function of angle $90^{\circ}-i$ for an ISCO ($a=0$) orbit at $R=6\, R_g$ (blue) and an orbit at $R=10\, R_g$ (red), each for two spectral indices, $\alpha=0.6$ (solid) and $\alpha=1.5$ (dotted). A boosting factor $<1.5$ plausibly inferred from the light curves in Figures \ref{fig:Fig1}, \ref{fig:FigB2}, and  \ref{fig:FigB3} suggests $i<27^{\circ}$. At such low inclinations, almost face on, one would also expect little lensing and multiple images for $R\ge 6 \,R_g$ (see \citealt{broderick2006, hamaus2009}), again consistent with our observations. We have found a consistent inclination constraint using the flux modulation in light curves from the relativistic hot spot models described above.

 \section{Polarization Loops} 
\label{sec:AppD}

\begin{figure*}[ht]
\includegraphics[width=18cm]{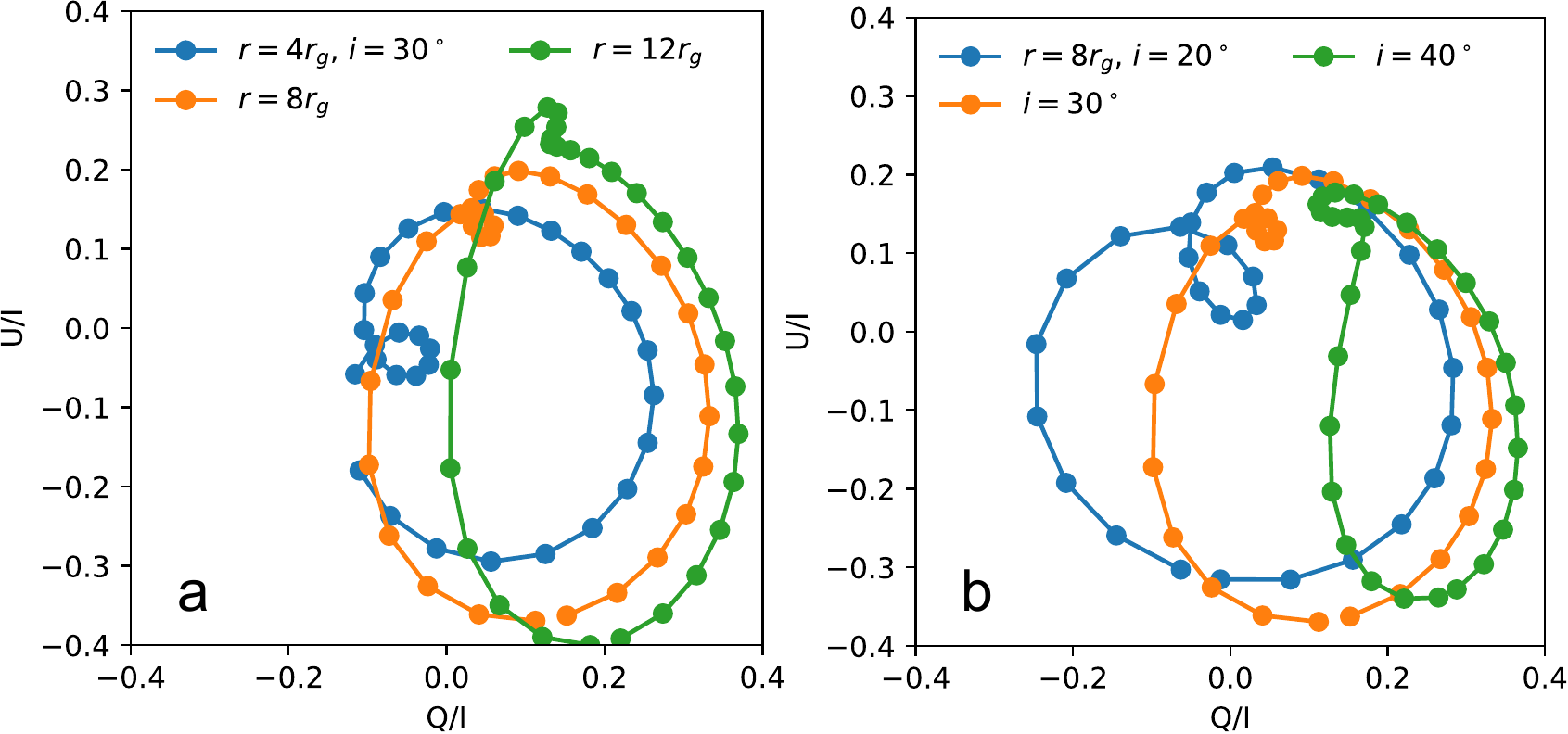}
\caption{Left (a): Simulation of the motion of a polarized, orbiting hot spot for a poloidal magnetic field configuration, for 30 degrees inclination and three different orbital radii $R = 4,8,12 \,R_g$ (b) and $R = 8 \,R_g$ at varying inclination. The observed July 28 flare polarization data are consistent with the Q/U loop signature and period for the $8\, R_g$, $i = 30^{\circ}$ curve (but at lower polarization fraction).}
\label{fig:FigD1}
\end{figure*}

During the July 28 flare we observed both polarization components with the half wave plate, as discussed in the instrumental description above. Figure \ref{fig:Fig2} (d) shows the location of the July 28$^{\rm }$ flare in the Q (horizontal) and U (vertical) plane, normalized to the total intensity and with a typical rms error shown in the lower left. The fractional polarization is 20-40\%, consistent with past single telescope results \citep{trippe2007,eckart2008,shahzamanian2015}. During the 100-minute duration of the flare the Q-U Stokes components trace out about 1.3 times a full "loop" or $180^{\circ}$ rotation of the polarization angle. We estimate the period for the polarization loop to be $P_{\rm pol}\sim48\pm6$ minutes, comparable to the orbital period of the centroid motion in that flare (Figure \ref{fig:Fig3}). For the May 27$^{\rm }$ and July 22$^{\rm }$ flares we only recorded the Q-polarization. Figure \ref{fig:Fig2} (a) shows that Q varied over a full period during the $\sim$70-minute duration of the May 27 flare, again comparable to the orbital period, while the polarization on July 22 rises from zero to a peak and goes back to zero, consistent with half of a period.\\
The high Sgr A* near-IR polarization arises from optically thin synchrotron emission from energetic electrons (energy $\gamma =\frac{E}{m_{e} c^2}\approx 10^3$). In the hotspot model described above, the polarization angle as a function of time traces out the magnetic field geometry in the accreting gas around the black hole. If the field is initially weak, it will be sheared into a toroidal configuration \citep{balbus1991} while any accreted magnetic flux will form a magnetically dominated atmosphere near the pole. If sufficient flux can be advected inwards, the fields can become dynamically important with predominantly vertical magnetic field at all latitudes near the black hole \citep{narayan2003}. 
Figure \ref{fig:FigD1} shows models of the motion of the Q-U Stokes parameters at three different radii on a circular orbit around a Schwarzschild black hole, where the dominant magnetic field direction was poloidal, or perpendicular to the orbital plane. In this case the polarization angle varies in loops with a period $P_{\rm pol}=P_{\rm orbit}$, as in the data of Figure \ref{fig:Fig3}. The loops are more pronounced and centered on (Q,U)=(0,0) for smaller orbital radii. In contrast, for a toroidal field configuration, one would expect loops with half the orbital period, $P_{\rm pol}=0.5\times P_{\rm orbit}$, or two loops per orbital period \citep{bromley2001,dexter2016}. For the range of inferred radii from the astrometry data, producing a loop passing through (0,0) as observed requires an inclination of $i\sim15-30^{\circ}$. This low inclination is consistent with both the apparent orbit on the sky and the lack of strong flux modulation from Doppler beaming or gravitational lensing.\\
The loops in the poloidal field case are caused by light bending, which adds an inwards radial component to the wave vector and an azimuthal component to the polarization map. At low inclination and small radius this effect becomes strong enough to produce one large loop per orbit. The observed polarization signature is therefore further evidence that the material is orbiting close to the black hole. Our finding of a likely poloidal field geometry further suggests that the Sgr A* accretion zone may be magnetically dominated, consistent with the high field strength inferred at $\sim0.1$\,pc from the rotation measure of the magnetar SGR J1745-2900 \citep{eatough2013}. \\Alternatively, such a field configuration could arise in a jet. If part of a collimated outflow, the emission radius would need to be substantially larger than what we observe and be only mildly relativistic to obey our observed limits on Doppler beaming and apparent out-of-plane motion.

\end{appendix}

\end{document}